# General Synthetic Route Towards Highly Dispersed Metal Clusters Enabled by Poly(ionic liquid)s

Jian-Ke Sun,[†] Zdravko Kochovski,[‡] Wei-Yi Zhang,[†] Holm Kirmse,[§] Yan Lu,[‡,∥] Markus Antonietti,[†] and Jiayin Yuan[*,†,⊥]

[†]Max Planck Institute of Colloids and Interfaces, Department of Colloid Chemistry, D-14424 Potsdam, Germany

[‡]Soft Matter and Functional Materials, Helmholtz-Zentrum Berlin für Materialien und Energie, Hahn-Meitner-Platz 1, 14109 Berlin, Germany

[§]Institute of Physics, Humboldt University of Berlin, D-12489 Berlin, Germany

[∥]Institute of Chemistry, University of Potsdam, 14476 Potsdam, Germany

[⊥]Department of Chemistry and Biomolecular Science & Center for Advanced Materials Processing, Clarkson University, 8 Clarkson Avenue, Potsdam, New York 13699, United States

Supporting Information Placeholder

**ABSTRACT:** The capability to synthesize a broad spectrum of metal clusters (MCs) with their size controllable in a sub-nanometer scale presents an enticing prospect for exploring nanosize-dependent properties. Here we report an innovative design of a capping agent from a polytriazolium poly(ionic liquid) (PIL) in a vesicular form in solution that allows for crafting a variety of MCs including transition metals, noble metals, and their bimetallic alloy with precisely controlled sizes (~ 1 nm) and record-high catalytic performance. The ultrastrong stabilization power is result of an unusual synergy between the conventional binding sites in the heterocyclic cations in PIL, and an *in-situ* generated polycarbene structure induced simultaneously to the reduction reaction.

## INTRODUCTION

Metal clusters (MCs) with dimensions between a single metal atom and nanoparticles of > 2 nm have attracted focused attention.[1] The small size of MCs forces most of, if not all, their constitutional atoms to be exposed to surface. In addition, the size approaches the Fermi wavelength of electrons, resulting in molecule-like characters including discrete energy levels, size-dependent fluorescence, good photostability and in some cases biocompatibility.[2] However, the extremely tiny size brings synthetic difficulty because such clusters are naturally prone to aggregation and particle growth, driven by high surface energy.[3] Some synthetic routes have been developed over the past few years to prepare MCs.[4] As a popular one, an impregnation-precipitation procedure to physically confine MCs inside a nanoporous support has been described,[5] but the precise size control of uniformity of tiny pores of the support itself is challenging, not to mention the tedious surface functionalization and retarded diffusion kinetics in such systems. A solution to this problem is to apply, instead of a solid porous support, capping agents, often in large excess, targeting a maximum surface coverage to lower down surface energy and suppress growth.[6] This simple and scalable method however suffers frequently from a "trade-off", that is covering a high-energy surface which is catalytically or photochemically active adversely restricts accessibility to active sites. Generally speaking, a "smarter" design of capping agents is constantly pursued to reach an optimal balance or even break such trade-off.

Poly(ionic liquid)s (PILs) are an emerging, powerful class of functional polymers prepared from ionic liquids.[7] They possess a wide property and application spectrum due to the structural synergy between the ionic liquid component and the macromolecular architecture.[8] One particular feature of PILs is their unusual interfacial activity to bind eventually to most surfaces, from bio-nanomaterials, metals to carbons, serving as a type of universal stabilizer.[9] Recently, PILs have also contributed to nanomaterial design by generating unusual inorganic morphologies inaccessible so far in PIL-free synthetic conditions.[10] Herein, we report a further contribution by the construction of high-performance capping agents derived from PILs that can stabilize a bunch of well-dispersed, long-term stable MCs of extremely small size (~1 nm) and high catalytic performance. This success relies on the association of the traditional binding power of PILs to metal species, as mentioned previously, with a finely tuned hydrophilicity/hydrophobicity balance, but more significantly the simple in-situ formation of a poly(N-heterocylic carbene) (polyNHC) structure from the polytriazolium chains during the MC formation.

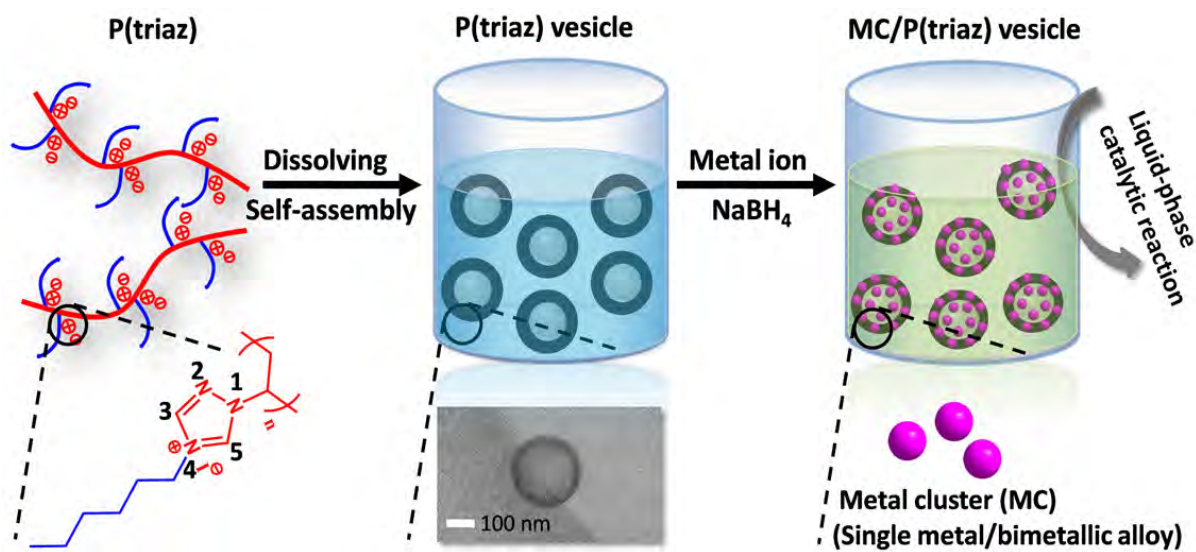

**Scheme 1.** Schematic illustration of the general procedure toward the synthesis of metal clusters (MCs) stabilized by structurally well-defined P(triaz) (the positions of atoms in the triazolium ring are numerically labeled) vesicles in a dichloromethane and methanol mixture (volume ratio = 2:1). The inset at the bottom middle is a cryo-EM image of a single vesicle. See Figure S9 for an overview.

## RESULTS AND DISCUSSION

We started by synthesis of a 1,2,4-triazolium PIL poly(4-hexyl-1-vinyl-1,2,4-triazolium iodide) (denoted as "P(triaz)") with pending hexyl substituent along its polytriazolium backbone (Scheme 1 and S1-4). The hydrophilic triazolium iodide ion pair enables the P(triaz) well-soluble in polar solvents such as alcohols, while the pending hexyl chain expands their solubility/dispersability window to moderately polar solvents such as dichloromethane.[11] The incorporation of these two components of distinctively different features into one repeating unit favors the formation of superstructures in solutions of selective polarity, as discussed later. The triazolium units along the backbone are a key structure motif here, as they will serve as precursors to polyNHC that amplifies their capping power. Compared with common elemental binding sites, *e.g.*, N, O and S, NHCs as strong σ-donators and comparatively weak π-acceptors are more robust to bind metal centers either homogeneously in solution or heterogeneously on metal nanoparticles.[12,13] Based on these considerations, the reaction medium used here is specifically chosen as a dichloromethane and methanol mixture (volume ratio = 2:1), in which P(triaz) self-assembles into vesicular structures (inset cryogenic electron microscopy (cryo-EM) images in scheme 1, Figure S9) of 40 ~ 250 nm in size and 15 ~ 35 nm in the wall thickness. The vesicular structures are, in comparison to classic polymer core-shell micelles, a favorable support for heterogeneous catalysis, as its thin wall facilitates fast mass diffusion and exposes its both surfaces to contact reactants in solution. Decreasing the alkyl length from hexyl to butyl (Figure S7) results in a molecularly dissolved PIL in the same mixture solvent, while a PIL with longer alkyl chain, such as decyl (Figure S8), forms irregularly shaped large colloids.

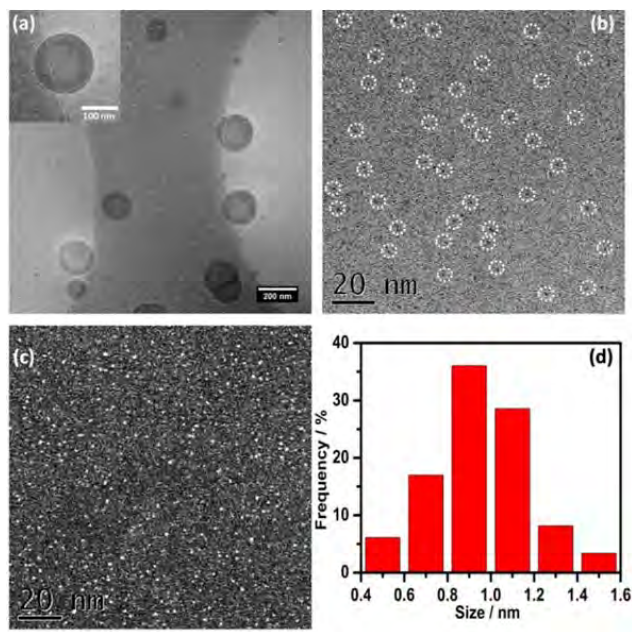

**Figure 1.** (a) Cryo-EM image of the Pd/P(triaz) hybrid vesicles (dark circle) on a Quantifoil carbon grid. The inset is a cryo-EM image of a single Pd/P(triaz) vesicle. (b) Bright field (BF) TEM image of Pd clusters on a P(triaz) vesicle. Some of Pd clusters are highlighted by white dotted circles. (c) HAADF-STEM image and (d) the corresponding size distribution histogram of Pd clusters.

In a first step, Pd was chosen as an example to demonstrate the effectiveness of our polymer support towards the synthesis of a wide spectrum of high-quality MCs (Scheme 1). Briefly, a light yellow solution of P(triaz) (5 mg) in a 9 mL mixture solution of dichloromethane and methanol (volume ratio = 2:1) was prepared. The color of solution became dark brown right upon adding palladium nitrate (Pd content in solution = 0.5 mg) (Figure S10), suggesting coordination bonding between Pd (II) and P(triaz). After aging for 20 min, the mixture was subsequently reduced by a methanol solution (0.5 mL) of sodium borohydride (NaBH4, 5 mg) to a light brown solution without any precipitate, indicating Pd cluster formation via reduction and stabilization by P(triaz).

The size and size distribution of Pd clusters were characterized by transmission electron microscopy (TEM) and cryo-EM. The cryo-EM and bright field (BF) TEM images of Pd/P(triaz) show the formation of well-dispersed P(triaz) vesicles and Pd clusters stabilized onto them, respectively (Figure 1a,b and S11&12). The morphology of Pd clusters was characterized by high-angle annular dark-field scanning transmission electron microscopy (HAADF-STEM) (Figure 1c). The average diameter of Pd clusters is 1.0 ± 0.2 nm (Figure 1d). X-ray photoelectron spectroscopy (XPS) analysis identifies the formed Pd clusters with binding energy at 335.3 and 340.6 eV (Figure S13), corresponding to Pd $3d_{5/2}$ and Pd $3d_{3/2}$ of metallic Pd, respectively. The variation of Pd loading concentration in the range of 4.8 ~14.5 wt% (with regard to the P(triaz)) did not change the ultrasmall size of clusters (~ 1 nm) (Figure S14). Despite the ultrasmall size, Pd clusters are uncommonly stable against aggregation upon storage of months. More attractively, the PIL-stabilized Pd clusters can be processed like polymers, *i.e.* solution-cast into a film on a substrate, dried and re-dissolved in solution (Figure S15). This is fundamentally different from common metal nanoparticles, which upon drying often break or become non-dispersible. In a control experiment, the P(triaz)-free metal ion solution upon reduction by NaBH4 contains seriously aggregating metal particles as revealed by TEM image in Figure S16.

To our surprise, the P(triaz) route described here is general towards a large variety of MCs, including transition (*e.g.*, Co, Ni, Cu) and noble metals (*e.g.*, Ru, Rh, Ag, Pt, Au,) and even their bimetallic alloy (*e.g.*, Au-Ni) with loading mass up to 29 wt%. These various types of MCs of averagely 1 nm in size are exceptionally dispersible in liquid-phase (Figure S10), and the representative HAADF-STEM images are collected in Figure 2 (Figure S17 with higher magnifications) and evidenced by XPS and powder X-ray diffraction (PXRD) characterizations (Figure S18-28).

The role of P(triaz) in stabilization and immobiliza-tion of MCs triggered our interest. The UV-Vis spectra were firstly recorded to monitor the solution mixing process of P(triaz) and metal ions. Notably, a shift of absorption band of the metal ion/P(triaz) mixture in comparison to individual metal ions or P(triaz) in solution was observed (Figure S29-36). For example, in the Pd(II)/P(triaz) system, compared to the absorption of Pd(II) ion at 209 nm and P(triaz) at 247 nm, new absorption bands at 231 and 295 nm emerged (indicated by black arrows in Figure 3b), indicative of strong affinity between both species in solution that enriches metal ions into the vesicular support. To specify the binding site of P(triaz), proton nuclear magnetic resonance ($^1$H NMR) measurements were employed. As shown in Figure 3c, upon addition of palladium nitrate the C5-proton signal in the triazolium ring (10.6 ppm) shifts to a high-field 10.3 ppm, indicative of its coordinative interplay with Pd(II). A similar trend was observed in other metal ion/P(triaz) mixtures as well (Figure S37&38). The critical step is the addition of NaBH$_4$ to the metal ion/P(triaz) solution. In Pd(II)/P(triaz), the C5-proton signal with a visible intensity drop could be observed (Figure 3c, highlighted by a light violet rectangle). The C5 protons in the triazolium ring are well-known to be active as carbene precursor even in the presence of weak bases.[13d] Our observation suggests that in the presence of $BH_4^-$ as a weak base that extracts protons (The chemical structure of P(triaz) maintained overall integrity in this case (see Figure S39) exception the extraction of part of its C-5 protons), carbenes form and bind the nearby metal species. Through integration of $^1H$ NMR spectra, 33 % of the triazolium units in P(triaz) were calculated to participate in carbene formation, corresponding to averagely one metal atom per 1 triazolium unit. Similar behaviors were found in other metal species with an estimated carbene/metal ratio between 1-2 (Table S1). The carbene-Pd complex formation was further confirmed by $^{13}C$ nuclear magnetic resonance ($^{13}C$ NMR) spectra (Figure 3d), in which a new peak appears at 169.6 ppm, typical for metal-carbene coordination.[14] Such new characteristic peaks assigned to metal-carbene complexes were also observed in other MCs/P(triaz) hyrbids (Figure S40). More evidences from the peak shift in XPS spectra of carbons after metal-carbene complex formation were collected in Figure 3e and Figure S41-48. For example, the C5 signal in carbene-Pd complex shifted 0.5 eV to lower binding energy due to the more pronounced nucleophilic characteristic (Figure 3e); this is consistent with previous observation in ionic liquid derived monocarbene to stabilize metal nanoparticles.[13f]

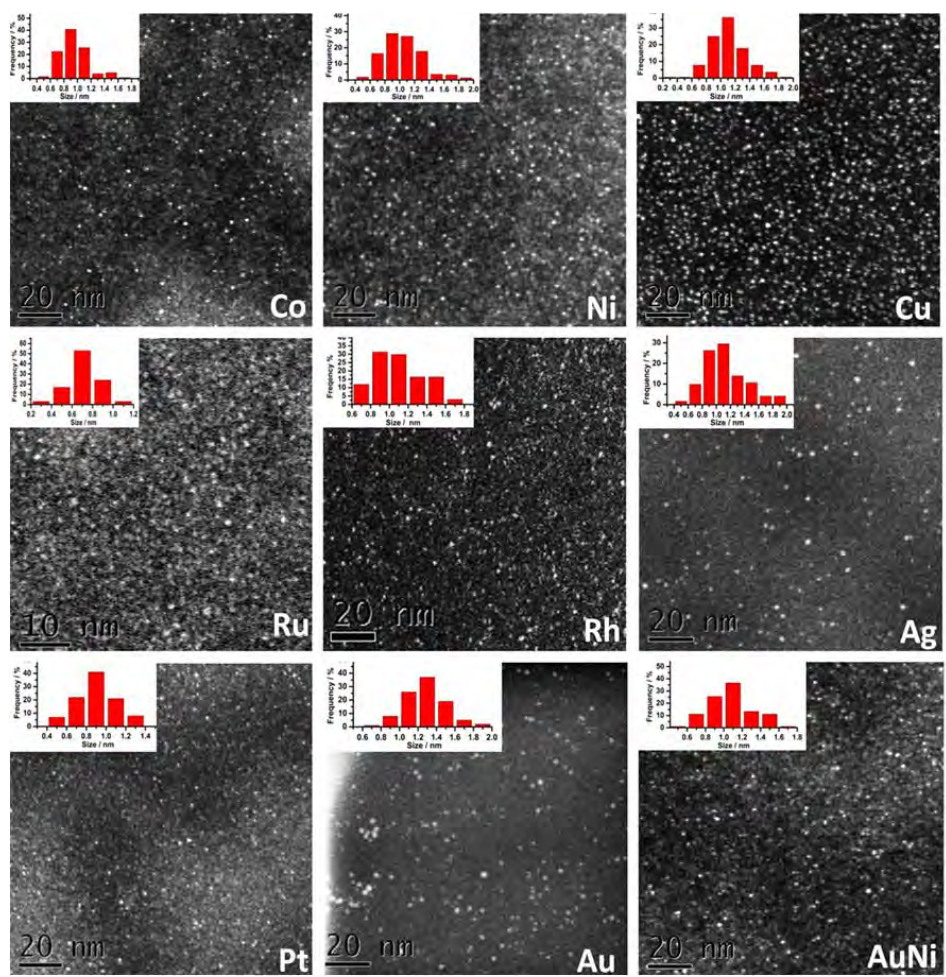

**Figure 2.** HAADF-STEM images of various MC/P(triaz) hybrids. Insets are the size distribution histograms of the corresponding MCs. The average size for Co: 0.9 ± 0.2 nm, Ni: 1.0± 0.2 nm, Cu: 1.1 ± 0.2 nm, Ru: 0.7 ± 0.2 nm, Rh: 1.1 ± 0.2 nm, Ag: 1.1 ± 0.2 nm, Pt: 0.9 ± 0.2 nm, Au: 1.2 ± 0.2 nm, and AuNi: 1.1 ± 0.2 nm..

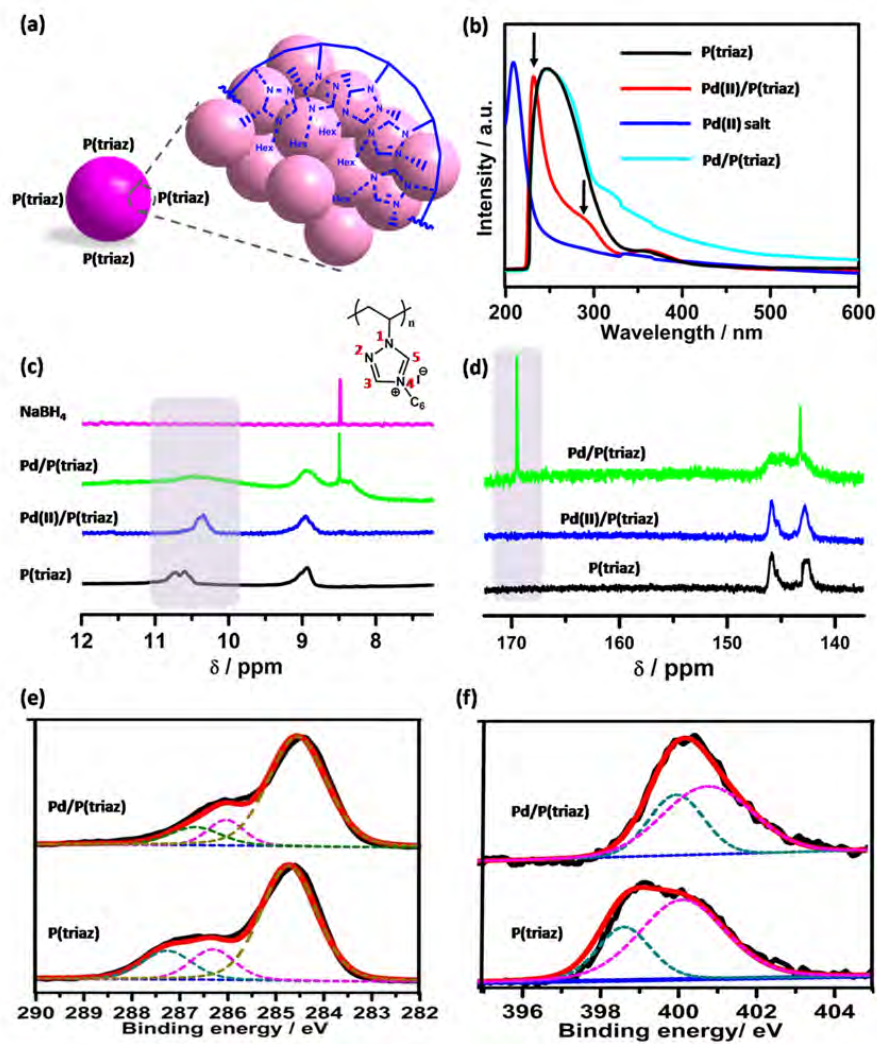

**Figure 3**. (a) Schematic illustration of a Pd cluster (deep violet ball) stabilized by P(triaz) with multi-binding sites, i.e. polycarbene and naked nitrogen sites, the light violet ball represents the component of Pd atoms. (b) The UV-vis spectra monitoring the formation process of Pd/P(triaz) in CH$_2$Cl$_2$ and CH$_3$OH (volume ratio = 2:1). (c) The $^1$H NMR spectra record the formation process of Pd/P(triaz) in CD$_2$Cl$_2$ and CH$_3$OH (volume ratio = 2:1). The signal at 8.5 ppm is attributed to NaBH$_4$ as a blank in CD$_2$Cl$_2$ and CH$_3$OH mixture (volume ratio = 2:1). (d) $^{13}$C NMR spectra of P(triaz), Pd(II)/P(triaz) and Pd/P(triaz) in CD$_2$Cl$_2$ and CH$_3$OH mixture (volume ratio = 2:1). A strong intensive peak at 169.6 ppm appeared in Pd/P(triaz) (shown in a light violet rectangle), which is a typical chemical shift for metal-carbene coordination that has been observed previously. (e) XPS spectra for C1s signals of P(triaz) and Pd/P(triaz). In P(triaz), the C1s spectra could be fitted by the sum of three separated peaks (dotted lines) with 1:1:8 area ratios that correspond to C5 (287.2 eV), C3 (286.3 eV) and eight alkane carbons (284.6 eV) in PIL (the charge of alkane carbons here were corrected to 284.6 eV), respectively. The C5 component in Pd/P(triaz) shifts 0.5 eV to lower binding energy (286.7 eV) as compared with that of P(triaz) due to the Pd-carbene complexation. (f) XPS spectra for N1s signals of P(triaz) and Pd/P(triaz). In both spectra, the N1s spectra could be fitted by the sum of two separated peaks (dotted lines) with 1:2 area ratios that correspond to the naked nitrogen (N2) and the two nitrogen atoms (N1 and N4) of the triazolium ring.

It is clear that a strongly coordinating support is essential to control over the nucleation and growth of MCs.[15] In this context, native P(triaz) is first of all capable of coordinating with Pd(II) primarily through the naked N2 nitrogen atom. The XPS measurements detected a 1.2 eV shift of N2 signal from 398.6 eV in native P(triaz) to a high binding energy position (399.8 eV) in the Pd/P(triaz) hybrid (Figure 3f), which is far more stronger than the shifts of N1 and N4 signals (<0.5 eV). Similar phenomena of binding energy shifts have also been observed in other MC/P(triaz) systems (Figure S49-56). The affinity of P(triaz) towards metal ions collects them inside the substrate and brings them closer to the triazolium ring for the subsequent metal-carbene complex generation. An extra function of N2 nitrogen is to facilitate the formation of carbene in triazolium compounds.[13d] Polycarbene formation by NaBH4 is a concurrent process to the reduction of Pd(II) and serves as in-situ capping agent for Pd clusters. Polycarbene is expected to act as a powerful complexing agent to transition metals, forms a thermodynamically more stable poly(carbene-metal) hybrid and plays a vital role in size control on MCs. It is different from previously reported soluble polymers, nanoporous materials or conventional poly(ionic liquid) supports, in which common elemental binding sites (e.g., N, O and S) are employed for metal nanoparticle stabilization.[5, 6, 9g,h, 17d] The lack of universal compatibility of these supports to different metals and their relatively mild coordination power make them difficult to apply to a broad spectrum of MCs and to effectively control their size and dispersability simultaneously. In a control experiment using poly[4-hexyl-1-vinylimidazolium iodide] (denoted as PIL-imidaz, Figure S5&6), in which the C2 position (N-CH2-N) is a weaker carbene precursor in the conditions used here for MC synthesis. The resultant Pd nanoparticles stabilized by PIL-imidaz are larger and broadly distributed in size from 1.5 to 11 nm (Figure S57). The $^1H$ NMR spectra confirmed that after addition of NaBH4, the C2-H signal has only a little shift (<0.2 ppm, Figure S58), indicating the C2-H in the imidazolium ring is more inactive for metal-carbene formation than that in triazolium. This is also quantified by our calculation from its $^1H$ NMR spectra that for the imidazolium units only less than 5 % in PIL-imidaz participate in the metal-carbene formation (Figure S59). It is clear that the in situ generated polycarbene is required for stabilizing ultrasmall MCs. This role can also not be replaced by analogous monocarbenes; by using the triazolium monomer instead of P(triaz) as stabilizer, the freshly prepared Pd particles aggregate macroscopically in solution (Figure S60). In fact, it was found that the C5-proton signal in P(triaz) has a slight decrease (< 8 %) in its integrity right upon Pd(II) addition, although the $^{13}$C-NMR spectrum detected no metal-carbene signal (Figure 3d). This evidence implies the replacement of the C5-proton by Pd(II) that forms metal-carbene complex might take place already before NaBH4 addition but in a weak equilibrium. This equilibrium shifts towards the metal-carbene side once the released C5-protons are neutralized by $BH_4^-$, which serves simultaneously as reductant and base. In a control experiment that replaced NaBH4 by UV light to reduce Ag(I)/P(triaz) to Ag/P(triaz), it ended up with large Ag aggregate (Figure 61&62), as metal-carbene formation is restricted in the absence of base.

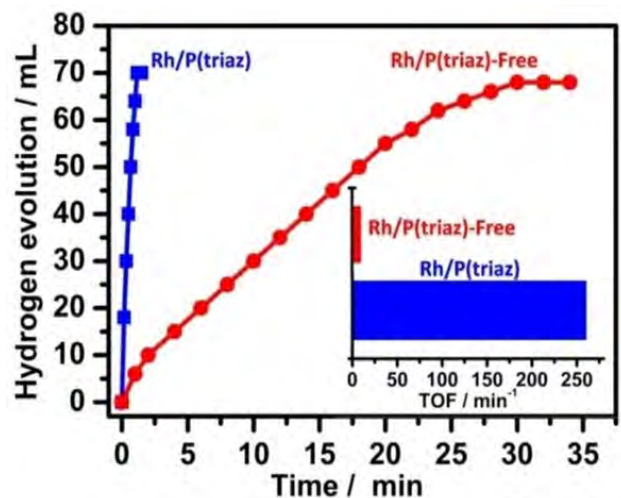

**Figure 4**. Time course plots of H2 generation for the methanolysis of ammonia borane (AB) over the Rh/P(triaz) and unsupported Rh-catalysts (Rh/P(triaz)-Free) at 298 K (Rh/AB = 0.01). Inset: the corresponding TOF values of the catalysts.

The P(triaz)-stabilized MC systems were exemplarily tested in a simple assay to evaluate the effect of the new capping agent on metal reactivity, the liquid-phase methanolysis of ammonia borane ($NH_3BH_3$, AB). AB is known as a chemical hydrogen source that releases hydrogen through reaction with water or methanol.[16] Rh clusters were chosen as catalyst, as they are commonly used for this reaction. Under identical experimental conditions, Rh/P(triaz) with a Rh loading of 0.01 mmol (Rh/AB = 0.01) completed the reaction (H2/AB = 3.0) within 1.16 min at 298 K (Figure 4), *i.e.* a turnover frequency (TOF) value as high as 260 min$^{-1}$. It is the highest value reported so far among all supported catalysts and capping agent protected catalysts,[16,17] including common commercial catalyst Pd/C (Figure S63). By contrast, the PIL-free Rh nanoparticles (4-8 nm in size) required more than 30 min (TOF number of 10 min$^{-1}$) to complete the reaction (Figure 4). In the case of triazolium monomer-stabilized Rh nanoparticles, the reaction ended up with a lower TOF number of 75 min$^{-1}$ (Figure S64). Even dendrimers (*e.g.*, polyamidoamine dendrimer with hydroxyl surface group, PAMAM-OH) as the classic high-performance stabilizer for Rh cluster (1.5 ± 0.3 nm, Figure S65) revealed a TOF number of "only" 141 min$^{-1}$ (Figure S66). Durability and recyclability of MCs are of high significance for potential catalytic applications. The Rh/P(triaz) catalyst was found to be effective for at least 4 successive additions of AB into the reactor (Figure S67). HAADF-STEM measurements reveal that the size and morphology of Rh cluster remained the same (Figure S68&69). The Rh/P(triaz) catalyst could be also recovered as a solid when dried in N2 atmosphere, washed and re-dissolved in solution for the next run without any activity loss (Figure S70). We assume that this unexpected high catalytic activity and stability can be reasonably attributed to the polycarbene-based capping agent that is used in this work to replace the polyimidazolium-based conventional PILs as well as the dendrimers. The metal-carbene bond might be strong enough that the metal-metal bonding is no more favored, *i.e.* from a thermodynamic perspective the MCs with surface metal-carbene complex are self-stabilizing. Besides, though the P(triaz) vesicle itself is catalytically inactive to AB (Figure S71), the vesicular form bearing a thin wall is a structural merit, as it enriches active sites necessary for MC stabilization and meanwhile facilities fast mass diffusion in solution. When using non-vesicular P(triaz) in methanol or PIL-butyl in the same dichloromethane/methanol mixture as stabilizers for Rh clusters (particle size: 1.4 ~ 1.8 nm), lower catalytic activity with a TOF number of 112 and 143 min$^{-1}$, respectively, was observed (Figure S72-75).

## CONCLUSION

In conclusion, polytriazolium PILs bearing hexyl side chains spontaneously assembled to nanovesicles and were capable of exerting a strict size control over a broad of MCs of about 1 nm in size. It prominently enhances catalytic performance, *e.g.* the reactivity of the as-synthesized Rh MC/P(triaz) catalyst is currently the best in a model reaction of methanolysis of ammonia borane, being at least 25% higher than the state-of-the-art catalyst.[17d] The key to the success of this approach is enhancing the already high stabilization power of PILs by in-situ generated polycarbenes under base addition. This previously undiscovered mechanism in MC stabilization is inspiring to researchers in nanoscience and particle processing. Beyond catalysis, the highly stable, dispersible MCs/PIL covering a broad spectrum of metal species are expected to be potentially also useful in imaging, nanomedicine and magnetism study.

## ASSOCIATED CONTENT

**Supporting Information**.
Supporting Information is available free of charge via the Internet at http://pubs.acs.org."
Experimental sections, Chemicals and instrumentations, additional characterizations, and catalytic data.

## AUTHOR INFORMATION

**Corresponding Author**
jiayin.yuan@mpikg.mpg.de.

**Notes**
The authors declare no competing financial interest.

## ACKNOWLEDGMENT

This work was supported by the Max Plank Society. J.Y. acknowledges the European Research Council Starting Grant (639720-NAPOLI) and a Startup Grant from Clarkson University. J. K. S thanks AvH (Alexander von Humboldt) foundation for a postdoctoral fellowship, and A. Voelkel for the AUC measurements. The authors also thank the Joint Lab for Structural Research at the Integrative Research Institute for the Sciences (IRIS Adlershof).

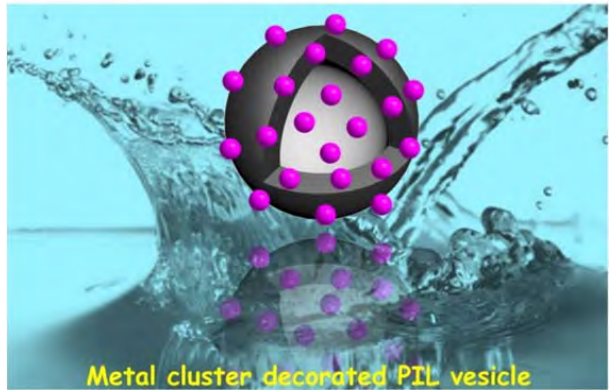

# Supporting Information

## 1. Chemicals and Instrumentation

All chemicals were from commercial sources and used without further purification.

Transmission electron microscopy (TEM) was performed on a JEOL 2010FS transmission electron microscope operated at 120 kV. Scanning transmission electron microscopy (STEM) was performed on a Jeol JEM-2200FS transmission electron microscope operated at 200 kV equipped with a high-angle annular dark-field (HAADF) STEM detector. Cryogenic transmission electron microscopy (cryo-EM) was performed with a JEOL JEM-2100 transmission electron microscope (JEOL GmbH, Eching, Germany). Cryo-EM specimens were prepared by applying a 4 μl drop of a dispersion sample to holey carbon-coated copper TEM grids (Quantifoil Micro Tools GmbH, Jena, Germany) and plunge-frozen into liquid ethane with an FEI vitrobot Mark IV set at 4°C and 95% humidity. Vitrified grids were either transferred directly to the microscope cryo transfer holder (Gatan 914, Gatan, Munich, Germany) or stored in liquid nitrogen. Imaging was carried out at temperatures around 90 K. The TEM was operated at an acceleration voltage of 200 kV, and a defocus of the objective lens of about 3.5 − 4 μm was used to increase the contrast. Cryo-EM micrographs were recorded at a number of magnifications with a bottom-mounted 4*4k CMOS camera (TemCam-F416, TVIPS, Gauting, Germany). The total electron dose in each micrograph was kept below 20 $e^-/Å^2$. Powder X-ray diffraction (PXRD) was carried out on an X-ray diffractometer of Rigaku, Ultima IV. X-ray photoelectron spectroscopy (XPS) studies were performed on a ThermoFisher ESCALAB250 X-ray photoelectron spectrometer (powered at 150 W) using Al K$_α$ radiation ($\lambda$ = 8.357 Å). To compensate for surface charging effects, all XPS spectra were referenced to the C 1s neutral carbon peak at 284.6 eV. The solution UV-Vis absorption measurements were recorded on a Lambda 900 spectrophotometer. $^1$H and $^{13}$C nuclear magnetic resonance ($^1$H-NMR) measurements were carried out at room temperature on a Bruker DPX-400 spectrometer in different deuterated solvents. The energy-dispersive X-ray (EDX) mapping measurements were taken on a Gemini scanning electron microscope (SEM) with an EDX spectrometer. Thermogravimetric analysis (TGA) experiments were carried out by a Netzsch TG209-F1 apparatus at a heating rate of 10 K min$^{-1}$ under a constant N$_2$ flow. Differential scanning calorimetry (DSC) measurements were conducted on a Perkin-Elmer DSC-1 instrument at a heating rate of 10 K min$^{-1}$ under a N$_2$ flow.

## 2. Experimental Section
### 2.1 Synthesis of poly(ionic liquid)s

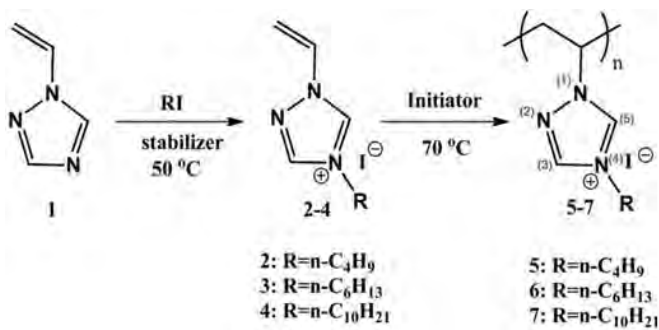

**Scheme S1.** Synthetic procedure of triazolium poly(ionic liquid)s with different alkyl chain length.

**Polymer 5**: Poly(4-butyl-1-vinyl-1,2,4-triazolium iodide) (simplified as "PIL-butyl"), the positions of atoms in the triazolium ring are highlighted in red color (similarly hereinafter).
**Polymer 6**: Poly(4-hexyl-1-vinyl-1,2,4-triazolium iodide) (simplified as "P(triaz)")
**Polymer 7**: Poly(4-decyl-1-vinyl-1,2,4-triazolium iodide) (simplified as "PIL-decyl")
**Monomer 2-4 synthesis**: A mixture of 1-vinyl-1,2,4-triazole **1** (5 mL, 5.5g, 57.83 mmol) and a 1.2 equivalent amount of n-iodoalkanes (1-iodobutane, 1-iodohexane and 1-iododecane) were added into a 100 mL round flask, accompanied with 2,6-di-tert-butyl-4-methylphenol (50 mg, 0.227 mol) as the stabilizer. After heated at 50 °C overnight, crude products were precipitated in diethyl ether and washed with the same solvent for three times. Pale yellow powders were obtained after purification.

**Polymer 5-7 synthesis**: A mixture of monomers (**2-4**) with AIBN (1.5 mol%) as initiator was added to anhydrous DMF (concentration: ~1 g monomer in 10 mL solvent) inside a 100 mL round-bottom schlenk flask. The flask was treated with three freeze-pump-thaw cycles and finally purged with argon. The reaction was stirred at 70 °C for 24 h under argon atmosphere. Yellow powders were obtained after dialysis against water and a vacuum drying process.

The preparation methods of 4-hexyl-1-vinyl-imidazolium iodide and poly(4-hexyl-1-vinyl-imidazolium iodide) (simplified as "PIL-imidaz") were described in the literature [Salamone, J. C.; Israel, S. C.; Taylor, P.; Snider, B., Synthesis and homopolymerization studies of vinylimidazolium salts. *Polymer,* 1973, *14*, 639-644.]

The chemical structures of the polymers used in the present work were confirmed by $^1$H NMR spectra in Figure S1-4.

**4-Butyl-1-vinyl-1,2,4-triazolium iodide** (**2**) (Yield: 92%, 14.84 g): $^1$H NMR (400 MHz, DMSO-$d_6$, δ, ppm): 10.55 (s, 1H), 9.45 (s, 1H), 7.50 (dd, 1H, $J_1$=16 Hz, $J_2$=8 Hz), 6.03 (d, 1H, $J$=16 Hz), 5.58 (d, 1H, $J$=8 Hz), 4.32 (t, 2H, $J$=8 Hz), 1.86 (m, 2H), 1.32 (m, 2H), 0.88 (t, 3H, $J$=8 Hz); $^{13}$C NMR (400 MHz, DMSO-$d_6$, δ, ppm): 145.25, 142.14, 129.75, 110.79, 48.18, 30.09, 19.27, 13.81.

**4-Hexyl-1-vinyl-1,2,4-triazolium iodide (3)** (Yield: 96%, 17.04 g): $^1$H NMR (400 MHz, DMSO-$d_6$, δ, ppm): 10.55 (s, 1H), 9.45 (s, 1H), 7.51 (dd, 1H, $J_1$=16 Hz, $J_2$=8 Hz), 6.04 (d, 1H, $J$=16 Hz), 5.58 (d, 1H, $J$=8 Hz), 4.31 (t, 2H, $J$=8 Hz), 1.88 (m, 2H), 1.26 (m, 6H), 0.83 (t, 3H, $J$=8 Hz); $^{13}$C NMR (400 MHz, DMSO-$d_6$, δ, ppm): 145.25, 142.15, 129.76, 110.72, 48.41, 31.00, 28.94, 25.58, 22.29, 14.31.

**4-Decyl-1-vinyl-1,2,4-triazolium iodide (4)** (Yield: 95%, 19.94 g): $^1$H NMR (400 MHz, DMSO-$d_6$, δ, ppm): 10.46 (s, 1H), 9.40 (s, 1H), 7.52 (dd, 1H, $J_1$=16 Hz, $J_2$=8 Hz), 6.05 (d, 1H, $J$=16 Hz), 5.58 (d, 1H, $J$=8 Hz), 4.29 (t, 2H, $J$=8 Hz), 1.87 (m, 2H), 1.28 (m, 14H), 0.84 (t, 3H, $J$=8 Hz); $^{13}$C NMR (400 MHz, DMSO-$d_6$, δ, ppm): 145.30, 142.19, 129.77, 110.64, 48.36, 31.74, 29.36, 29.24, 29.14, 29.03, 28.86, 25.93, 22.55, 14.41.

**Poly(4-butyl-1-vinyl-1,2,4-triazolium iodide) (PIL-butyl-C4) (5)** (Yield: 82%, 3.24 g): $^1$H NMR (400 MHz, DMSO-$d_6$, δ, ppm): 10.53 (br, 1H), 9.34 (m, 1H), 4.67 (m, 1H), 4.26 (br, 2H), 2.52 (br, 2H), 1.91 (br, 2H), 1.38 (m, 2H), 0.94 (br, 3H).

**Poly(4-hexyl-1-vinyl-1,2,4-triazolium iodide) (P(triaz)) (6)** (Yield: 76%, 2.78 g): $^1$H NMR (400 MHz, DMSO-$d_6$, δ, ppm): 10.51 (br, 1H), 9.33 (m, 1H), 4.75 (m, 1H), 4.23 (br, 2H), 2.58 (br, 2H), 1.90 (br, 2H), 1.30 (br, 6H), 0.85 (br, 3H).

**Poly(4-decyl-1-vinyl-1,2,4-triazolium iodide) (PIL-decyl) (7)** (Yield: 68%, 2.51 g): $^1$H NMR (400 MHz, DMSO-$d_6$, δ, ppm): 10.68 (br, 1H), 8.92 (br, 1H), 5.23 (m, 1H), 4.28 (br, 2H), 2.71 (br, 2H), 2.01 (br, 2H), 1.20 (br, 14H), 0.82 (br, 3H).

**2.2 Analytical Ultracentrifugation (AUC):** The weight-averaged molecular weight of P(triaz) and PIL-imidaz polymers were analyzed by AUC method. The sample was dissolved in ethanol with 0.5M NaI to avoid charge effects in the analysis. The partial specific volume of the samples was determined in a density oscillation tube (DMA 5000, Anton Paar, Graz) (0.673 ml/g for PIL-imidaz and 0.596 ml/g for P(triaz)) Equilibrium experiments have been performed on an Optima XLI ultracentrifuge (Beckman Coulter, Palo Alto) and interference optics. Seven concentrations have been analyzed at different speeds starting from 7500 rpm up to 30000 rpm. Data were evaluated with the program MSTAR (Kristian Schilling, Nanolytics, Germany).

**2.3 Synthesis of metal cluster (MC) stabilized by PILs**
**Synthesis of Pd/P(triaz).** In a typical synthesis, 9 mL of dichloromethane and methanol mixture (volume ratio = 2:1) containing 5 mg of P(triaz) was subsequently added to 0.5 mL of methanol containing Pd(NO$_3$)$_2$·2H$_2$O (0.5 mg Pd in content). The resultant mixture solution was further homogenized after aging for 20 min. Then, 0.5 mL of methanol solution containing 5 mg of NaBH$_4$ was immediately added into the above solution with vigorous shaking, resulting in a well transparent dispersion of Pd/P(triaz).

**Synthesis of other MCs stabilized by P(triaz).** The synthetic procedure used above to prepare Pd/P(triaz) was followed by using 0.5 mL of methanol containing AgNO$_3$, H$_2$PtCl$_4$, HAuCl$_4$·3H$_2$O, RuCl$_3$·3H$_2$O, Rh(OAc)$_3$, Ni(NO$_3$)$_2$·6H$_2$O, Co(NO$_3$)$_2$·6H$_2$O, or Cu(NO$_3$)$_2$·2.5H$_2$O (The metal in content is 0. 5 mg except for Ru, in which 1 mg metal was included) in place of Pd(NO$_3$)$_2$·2H$_2$O.

**Synthesis of Pd/PIL-imidaz.** The synthetic procedure used above to prepare Pd/P(triaz) was followed by using 9 mL of dichloromethane and methanol mixture (volume ratio = 2:1) containing poly(3-hexyl-1-vinylimidazolium iodide) (PIL-imidaz) (5 mg) in place of P(triaz).

**Synthesis of Pd/triazolium monomer.** The synthetic procedure used above to prepare Pd/P(triaz) was followed by using 9 mL of dichloromethane and methanol mixture (volume ratio = 2:1) containing triazolium monomer (5 mg) in place of P(triaz).

**Synthesis of Ag/P(triaz) by UV reduction (instead of NaBH$_4$ reduction).** 9 mL of dichloromethane and methanol mixture (volume ratio = 2:1) containing 5 mg of P(triaz) was subsequently added to 0.5 mL of methanol containing AgNO$_3$ (0.5 mg Ag in content). The resultant mixture was further homogenized after aging for 20 min. Then, solution was irradiated with UV light (125 mW cm$^{-2}$) for 6 h, resulting in Ag/P(triaz) nanoparticle.

## 2.4 Synthesis of catalyst for AB methanolysis reaction

**Synthesis of Rh/P(triaz) catalyst.** The synthetic procedure used above to prepare Pd/P(triaz) was followed by using 9 mL of dichloromethane and methanol mixture (volume ratio = 2:1) containing P(triaz) (2.5 mg), 0.5 mL of methanol containing Rh(OAc)$_3$, (1 mg Rh in content) in place of Pd(NO$_3$)$_2$·2H$_2$O.

**Synthesis of Rh/P(triaz) catalyst in pure methanol**. The synthetic procedure used above to prepare Rh/P(triaz) catalyst was followed by using 9 mL of methanol .

**Synthesis of Rh-P(triaz)-Free catalyst.** The synthetic procedure used above to prepare Rh/P(triaz) catalyst was followed by using 9 mL of dichloromethane and methanol mixture (volume ratio = 2:1) without P(triaz) as stabilizer.

**Synthesis of Rh/PIL-butyl catalyst**. The synthetic procedure used above to prepare Rh/P(triaz) catalyst was followed by using 9 mL of dichloromethane and methanol mixture (volume ratio = 2:1) containing the PIL-butyl (2.5 mg) in place of P(triaz).

**Synthesis of Rh-triazolium monomer catalyst.** The synthetic procedure used above to prepare Rh/P(triaz) catalyst was followed by using 9 mL of dichloromethane and methanol mixture (volume ratio = 2:1) containing the triazolium monomer (2.5 mg) in place of P(triaz).

**Synthesis of Ru/PAMAM-OH catalyst.** The synthetic procedure used above to prepare Rh/P(triaz) catalyst was followed by using 9 mL of dichloromethane and methanol mixture (volume ratio = 2:1) containing PAMAM-OH (2.5 mg) in place of P(triaz).

## 2.5 Catalytic activity characterization

**Procedure for the methanolysis of AB by Rh/P(triaz) catalyst:** The reaction apparatus for measuring the hydrogen evolution from the methanolysis of AB is as follows. In general, the as-synthesized Rh/P(triaz) catalyst was placed in a two-necked round-bottomed flask (30 mL), which was placed in a water bath under ambient atmosphere. A gas burette filled with water was connected to the reaction flask to measure the volume of hydrogen. The reaction started when AB (30.8 mg) in 0.8 mL methanol was added into the flask. The volume of the evolved hydrogen gas was monitored by recording the displacement of water in the gas burette. The reaction was completed when there was no more gas generation. The methanolysis of AB can be expressed as follows:

$$NH_3BH_3 + 4CH_3OH \rightarrow NH_4B(CH_3O)_4 + 3H_2 \quad (1)$$

**Procedures for the methanolysis of AB by Rh/P(triaz)-methanol, Rh/PIL-butyl, Rh-SP-Free, Rh-triazolium monomer catalyst and Ru/PAMAM-OH catalysts:** The procedures for the methanolysis of AB were similar to that of Rh/P(triaz) catalyst except different catalysts were used.

**Procedures for the methanolysis of AB by redissolving $N_2$ atmosphere dried Rh/P(triaz) powder in solvents.** The as-synthesized Rh/P(triaz) solution was first dried in $N_2$ atmosphere, which was further washed by water and dried in vacuum oven. The resultant powder was re-dissolved in 9 mL of dichloromethane and methanol mixture (2:1) to generate the solution catalysts for catalytic reaction.

## 3. Additional data and figures

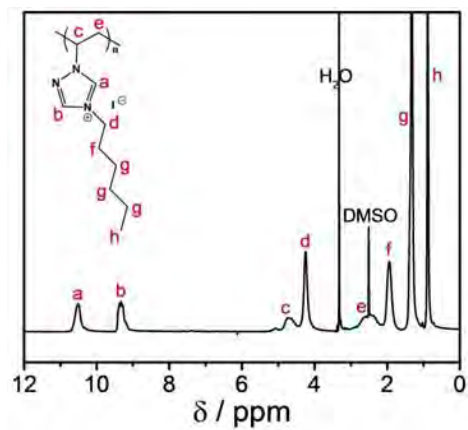

**Figure S1.** Chemical structure and $^1$H-NMR spectrum of poly(4-hexyl-1-vinyl-1,2,4-triazolium iodide) (P(triaz)).

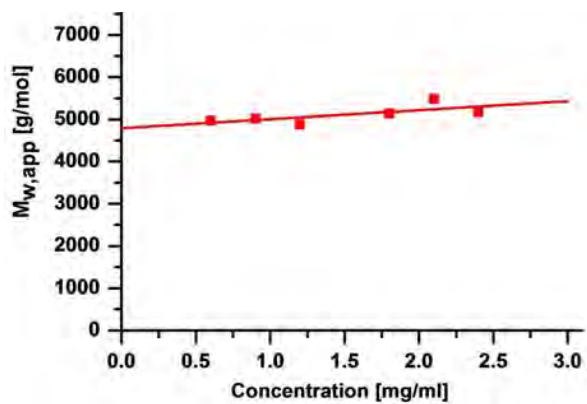

**Figure S2.** AUC analysis of P(triaz). The absolute weight-averaged molecular weight $M_w$ is determined to be $4.8 \times 10^3$ g mol$^{-1}$.

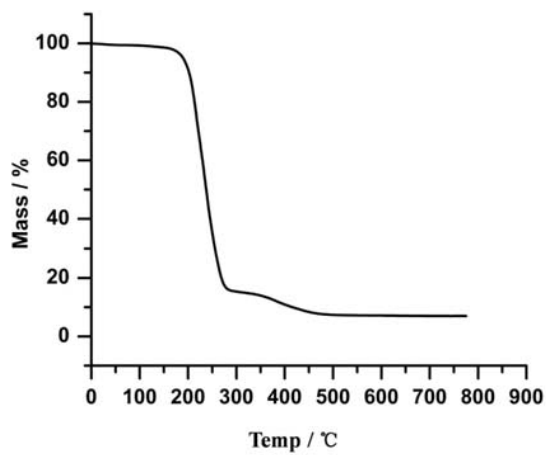

**Figure S3.** TGA plot of P(triaz) under $N_2$.

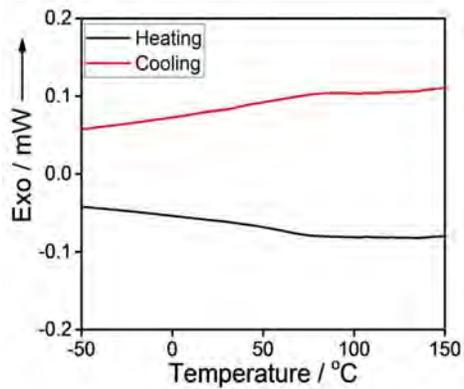

**Figure S4.** The DSC plot of P(triaz).

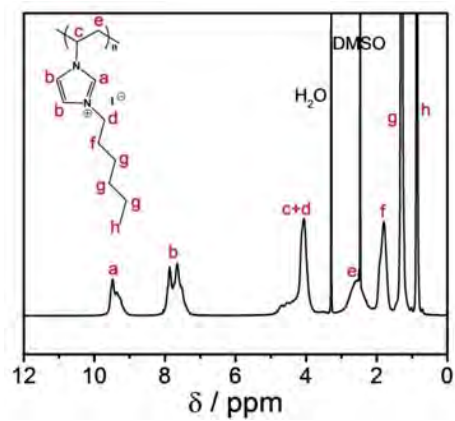

**Figure S5**. Chemical structure and $^1$H-NMR spectrum of poly(4-hexyl-1-vinyl-imidazolium iodide) (PIL-imidaz).

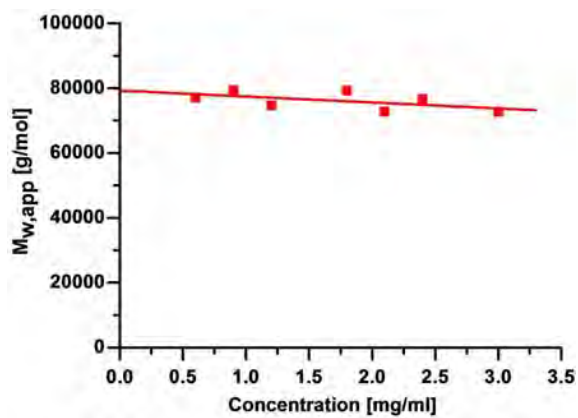

**Figure S6.** AUC analysis of PIL-imidaz. The absolute weight-averaged molecular weight $M_w$ is determined to be 7.9x10$^4$ g mol$^{-1}$.

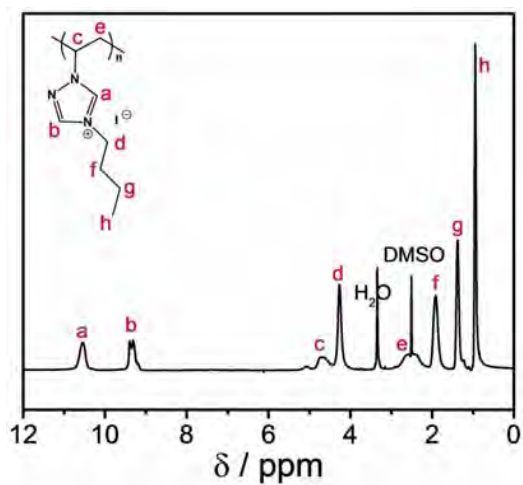

**Figure S7**. Chemical structure and ¹H-NMR spectrum of poly(4-butyl-1-vinyl-1,2,4-triazolium iodide) (PIL-butyl).

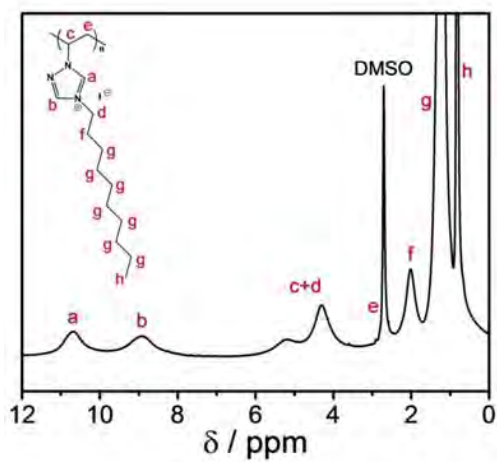

**Figure S8**. Chemical structure and ¹H-NMR spectrum of poly(4-decyl-1-vinyl-1,2,4-triazolium iodide) (PIL-decyl).

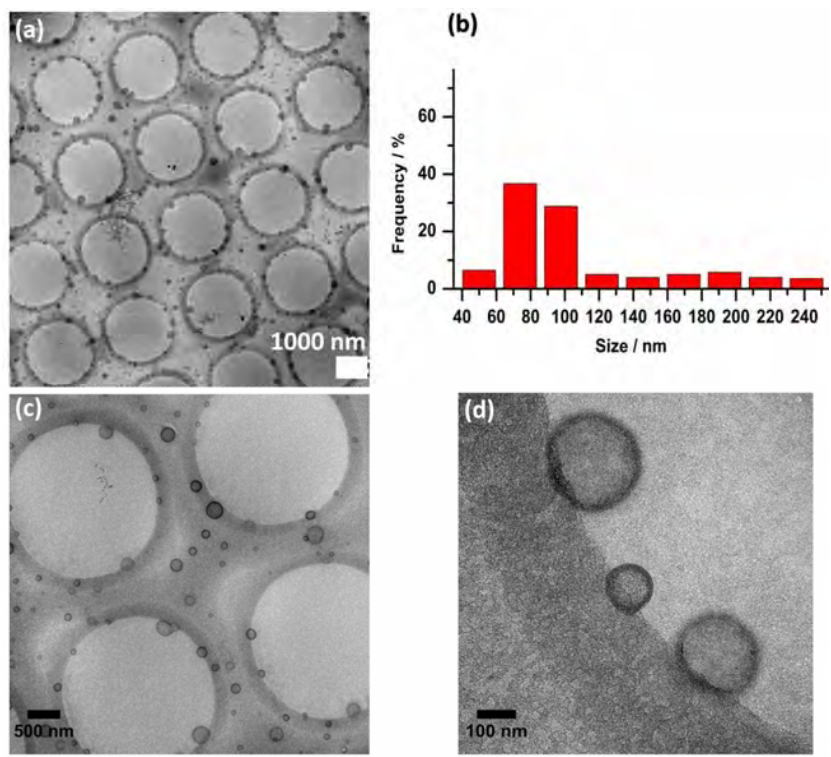

**Figure S9**. (a) A cryo-EM image of a low magnification of the vesicular P(triaz) in dichloromethane and methanol mixture (volume ratio = 2:1). Note, the bright large circles are from the Quantifoil carbon grid. (b) The corresponding size distribution histogram. (c-d) Cryo-EM images of vesicles at different magnifications.

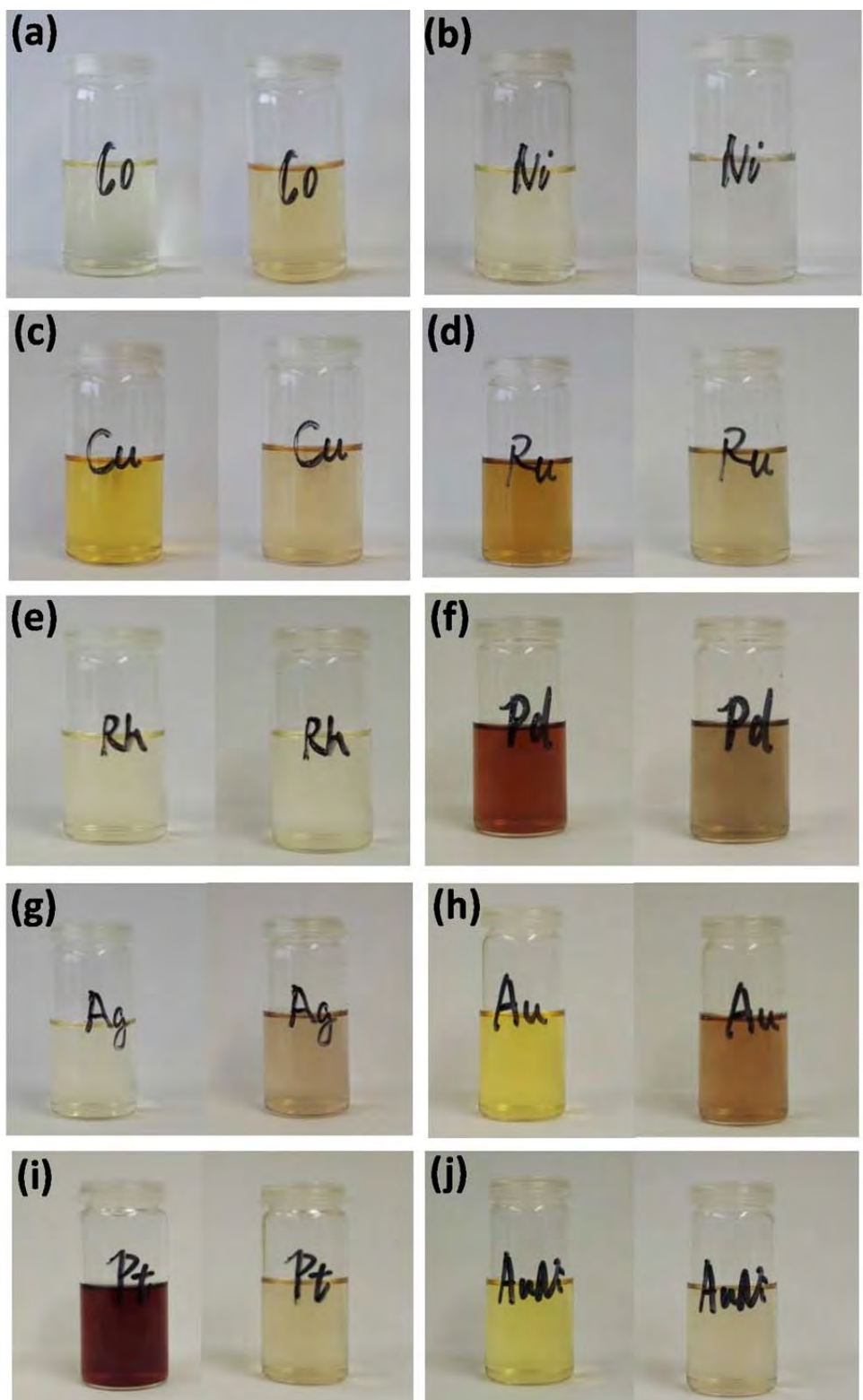

**Figure S10**. Photographs illustrating the synthesis of MC/P(triaz). The corresponding metal ion/P(triaz) (left) and MC/P(triaz) (right) in a dichloromethane and methanol mixture (volume ratio = 2:1) were shown in each photograph.

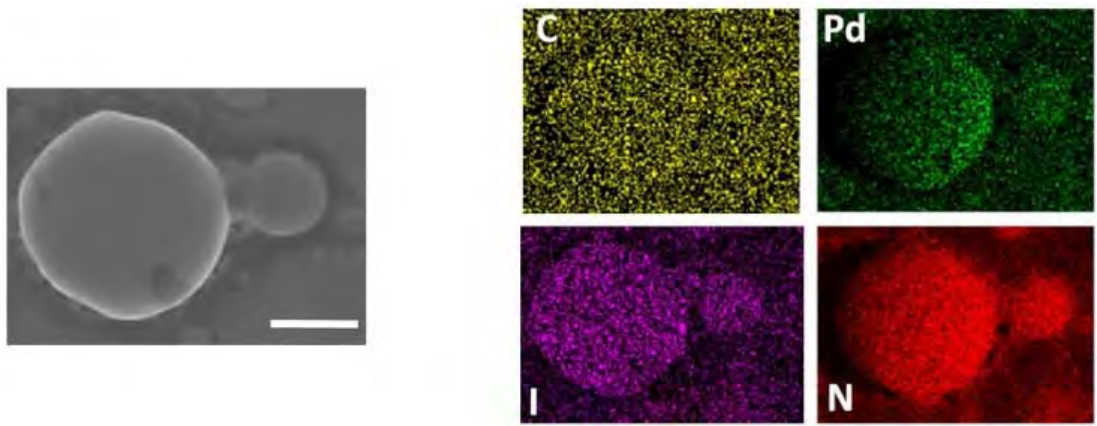

**Figure S11.** Left: a SEM image of a dried Pd/P(triaz) sample, scale bar: 5 μm (note that they are not the individual polymer vesicles but their aggregates due to the drying process). Right: The C, Pd, I and N elemental mapping of the same Pd/P(triaz) sample.

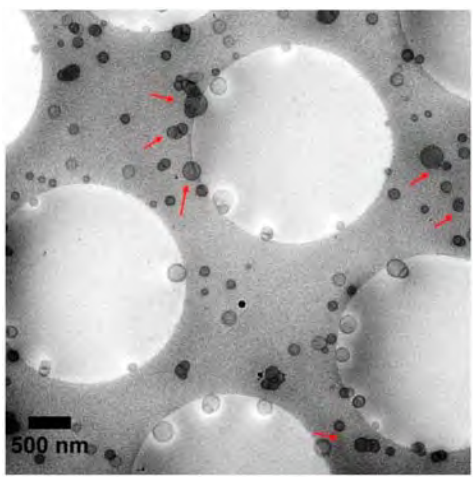

**Figure S12**. The cryo-EM image of the Pd/P(triaz) hybrid vesicles at a low magnification. Red arrows highlight the dark shell of the vesicles.

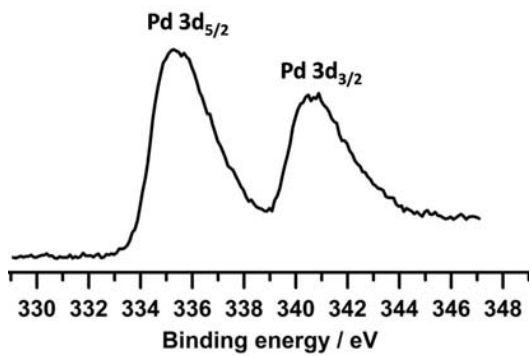

**Figure S13**. XPS spectrum of Pd/P(triaz) showing Pd $3d_{5/2}$ (335.3 eV) and $3d_{3/2}$ (340.6 eV) peaks of metallic Pd after Ar etching.

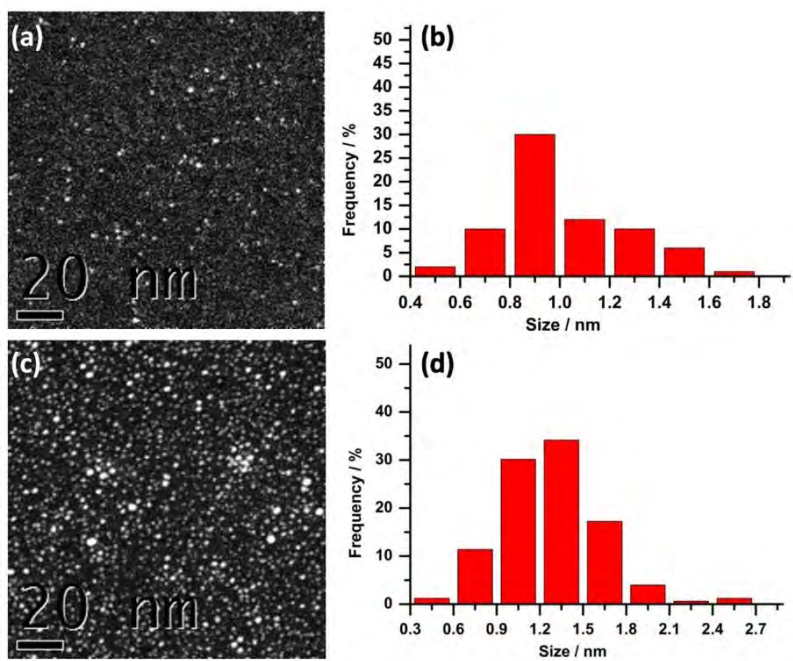

**Figure S14.** (a) HAADF-STEM image and (b) the corresponding size distribution histogram of Pd (4.8 wt%) clusters (size: 1.0 ± 0.2 nm). (c) HAADF-STEM image and (d) the corresponding size distribution histogram of Pd (14.5 wt%) clusters (size: 1.2 ± 0.3 nm).

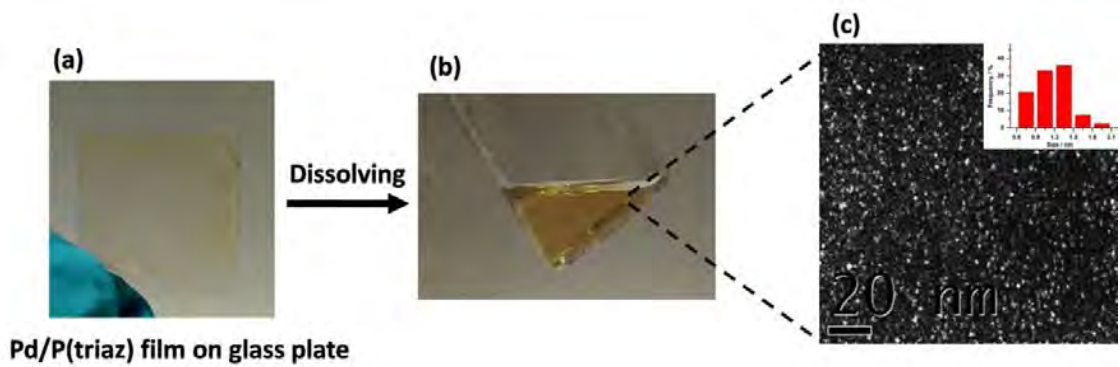

**Figure S15.** The illustration of the excellent processability of Pd/P(triaz) in a solution state. (a) The drop-casting of Pd/P(triaz) onto a piece of glass plate. (b) The film can re-dissolve in dichloromethane and methanol mixture (volume ratio = 2:1). (c) The HAADF-STEM image and the corresponding size distribution histogram (inset) of the redispersed Pd clusters.

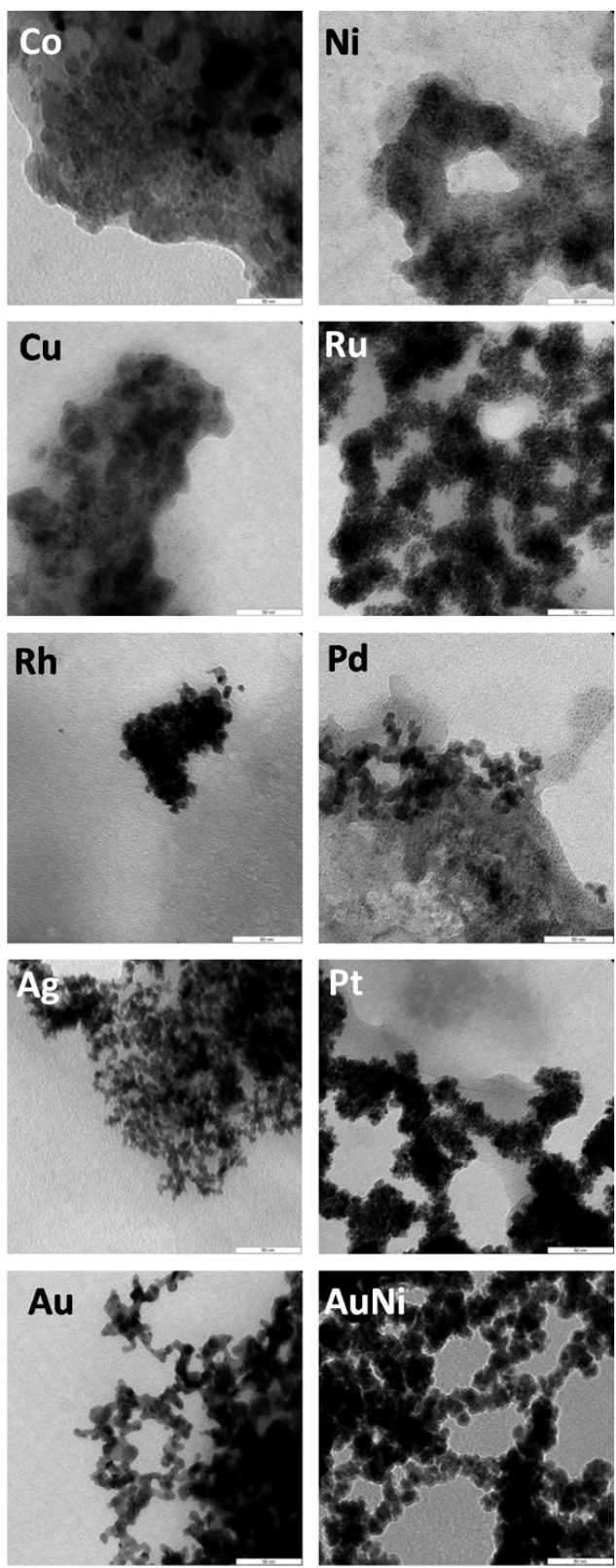

**Figure S16**. Bright field (BF) TEM images of different metal nanoparticles synthesized without P(triaz) stabilizer, scale bar, 50 nm.

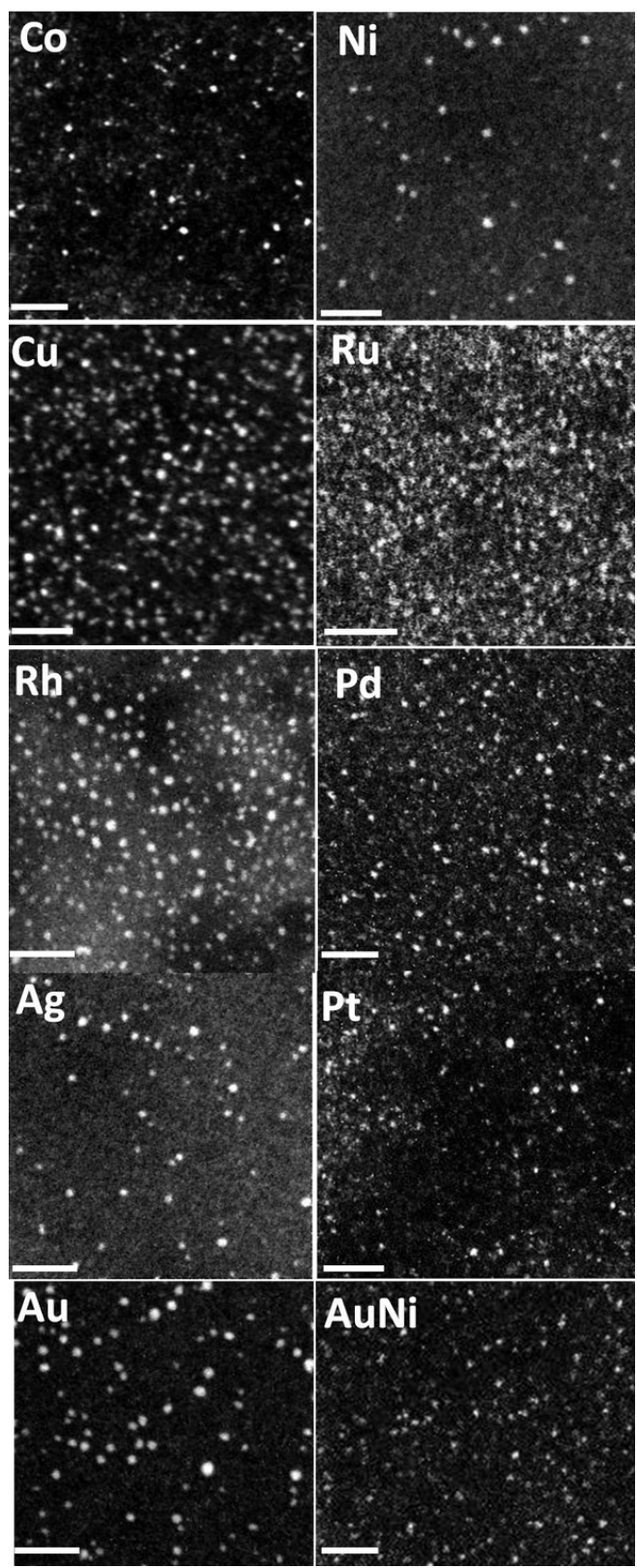

**Figure S17.** HAADF-STEM images of P(triaz) stabilized MCs with higher magnification, scale bar, 10 nm.

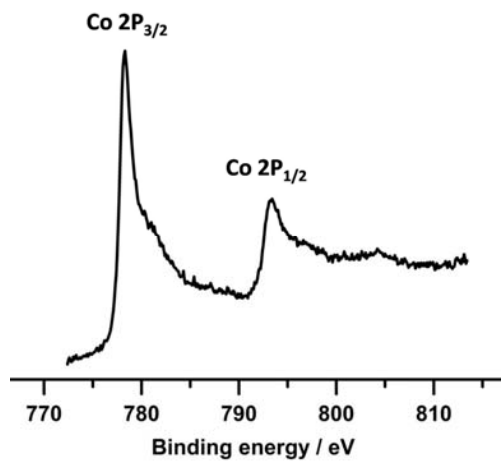

**Figure S18**. XPS spectrum of Co/P(triaz) showing Co 2p$_{3/2}$ (778.3 eV) and 2p$_{1/2}$ (793.3 eV) peaks of metallic Co after Ar etching.

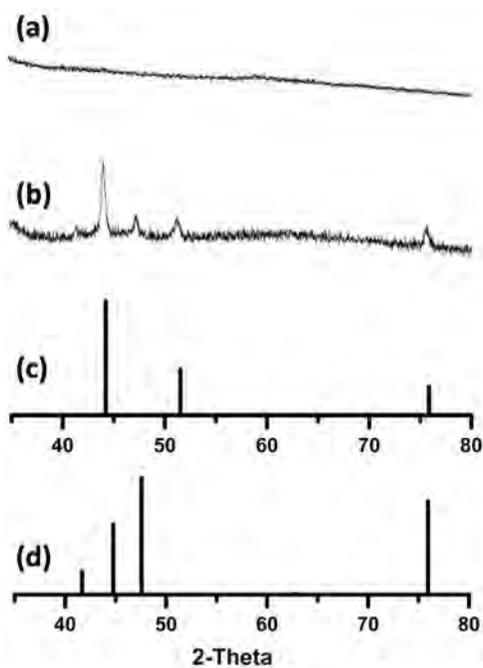

**Figure S19.** PXRD pattern of (a) the as-synthesized Co/P(triaz) and (b) after annealing at 500 °C for 3 h in Ar atmosphere. The patterns match well with (c) cubic Co (PDF#15-0806) and (d) hexagonal Co (PDF#05-0727). These experiments demonstrated that metallic Co is formed after reduction by NaBH$_4$.

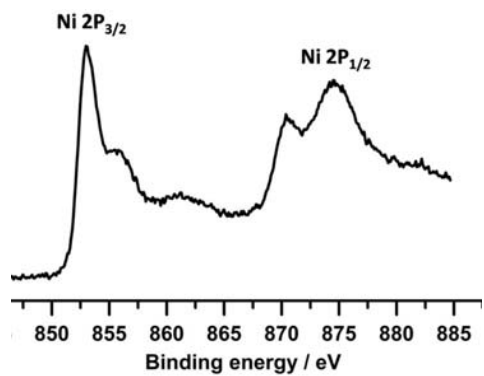

**Figure S20**. XPS spectrum of Ni/P(triaz) showing Ni 2p$_{3/2}$ (852.9 eV) and 2p$_{1/2}$ (874.5 eV) peaks of metallic Ni after Ar etching.

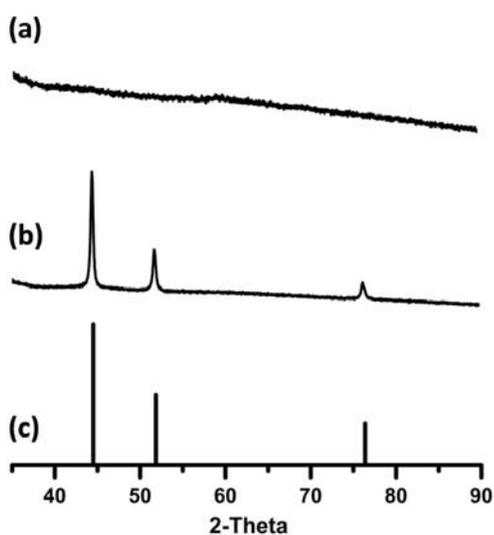

**Figure S21.** PXRD pattern of (a) the as-synthesized Ni/P(triaz) and (b) after annealing at 500 °C for 3 h in Ar atmosphere. The patterns match well with (c) cubic Ni (PDF#65-2865). These experiments demonstrated that metallic Ni is formed after reduction by NaBH$_4$.

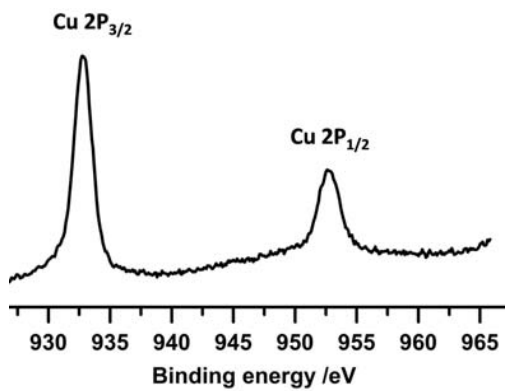

**Figure S22**. XPS spectrum of Cu/P(triaz) showing Cu 2p$_{3/2}$ (932.8 eV) and 2p$_{1/2}$ (952.6 eV) peaks of metallic Cu after Ar etching.

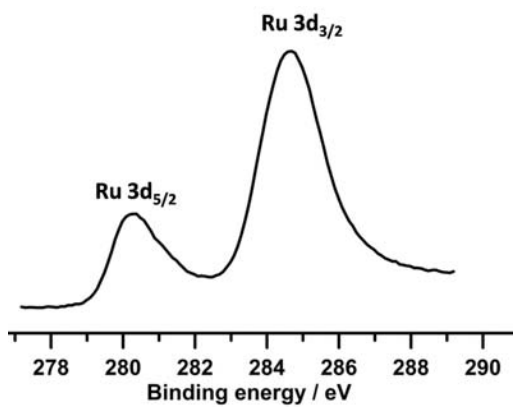

**Figure S23**. XPS spectrum of Ru/P(triaz) showing Ru $3d_{5/2}$ (280.2 eV) and $3d_{3/2}$ (284.6 eV) peaks of metallic Ru after Ar etching.

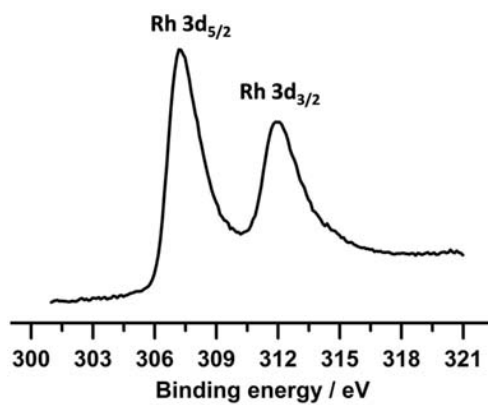

**Figure S24**. XPS spectrum of Rh/P(triaz) showing Rh $3d_{5/2}$ (307.2 eV) and $3d_{3/2}$ (311.9 eV) peaks of metallic Rh after Ar etching.

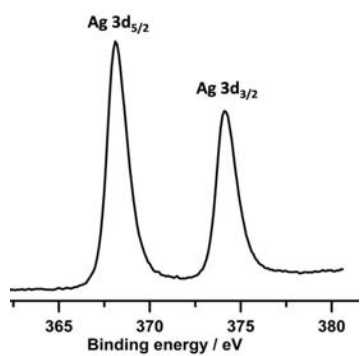

**Figure S25**. XPS spectrum of Ag/P(triaz) showing Ag $3d_{5/2}$ (368.1 eV) and $3d_{3/2}$ (374.1 eV) peaks of metallic Ag after Ar etching.

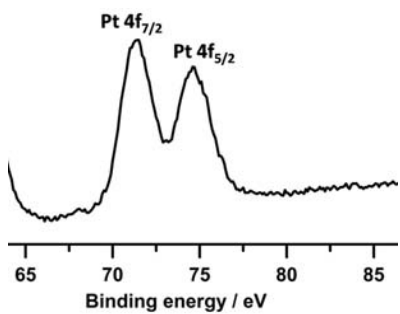

**Figure S26**. XPS spectrum of Pt/P(triaz) showing Pt 4f$_{7/2}$ (71.4 eV) and 4f$_{5/2}$ (74.6 eV) peaks of metallic Pt after Ar etching.

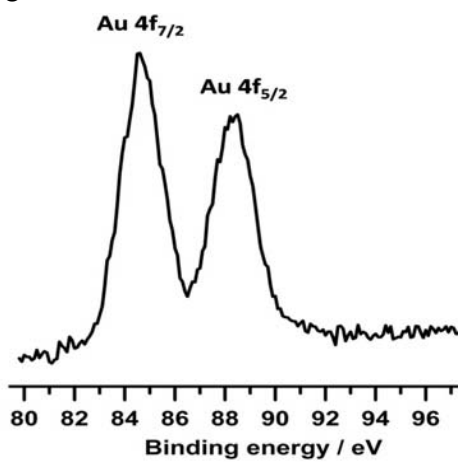

**Figure S27**. XPS spectrum of Au/P(triaz) showing Au 4f$_{7/2}$ (84.4 eV) and 4f$_{5/2}$ (88.3 eV) peaks of metallic Au after Ar etching.

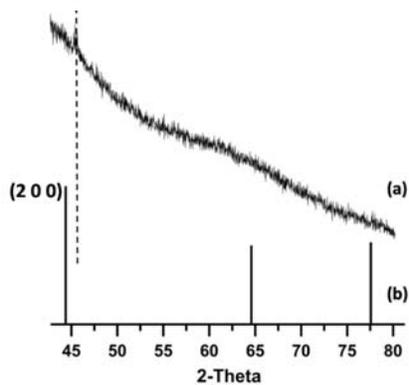

**Figure S28**. The PXRD patterns of (a) as-synthesized AuNi/P(triaz), (b) positions of reflections for pure Au marked by vertical lines. The (2 0 0) reflections of AuNi/P(triaz) shifted to higher angles (indicated by the dot line) as compared with that of pure Au. This observation indicates that Ni incorporated into the Au *fcc* structure forms an alloy phase.

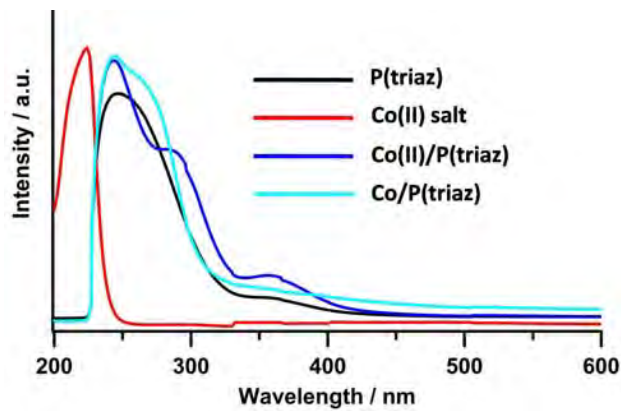

**Figure S29**. The UV-vis spectra monitoring the formation process of Co/P(triaz) in dichloromethane-methanol mixture (volume ratio = 2:1).

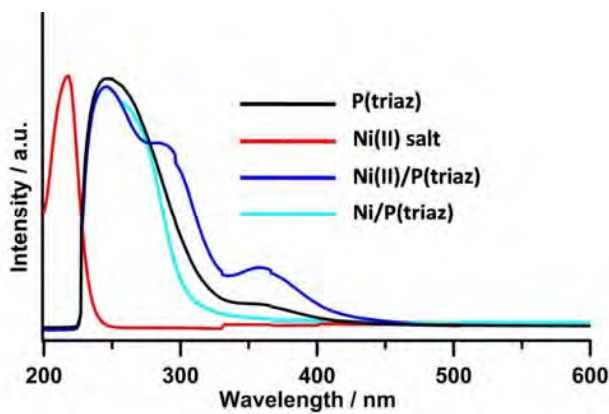

**Figure S30**. The UV-vis spectra monitoring the formation process of Ni/P(triaz) in dichloromethane-methanol mixture (volume ratio = 2:1).

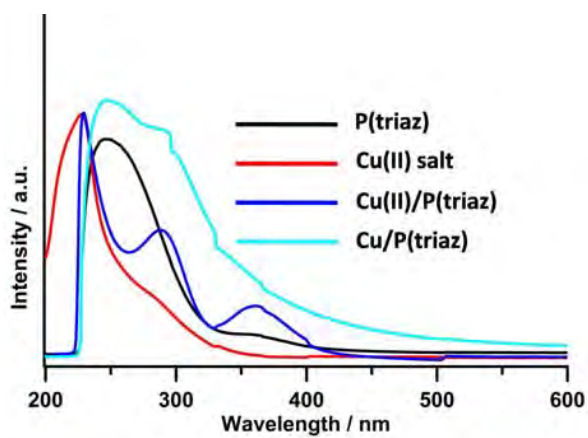

**Figure S31**. The UV-vis spectra monitoring the formation process of Cu/P(triaz) in dichloromethane-methanol mixture (volume ratio = 2:1).

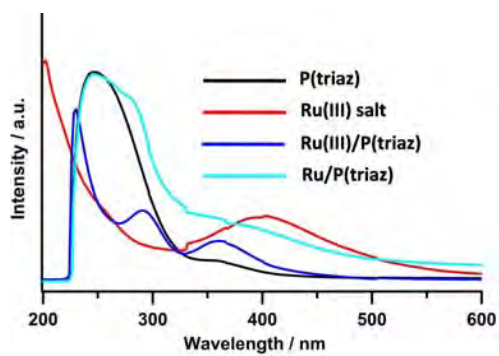

**Figure S32**. The UV-vis spectra monitoring the formation process of Ru/P(triaz) in dichloromethane-methanol mixture (volume ratio = 2:1).

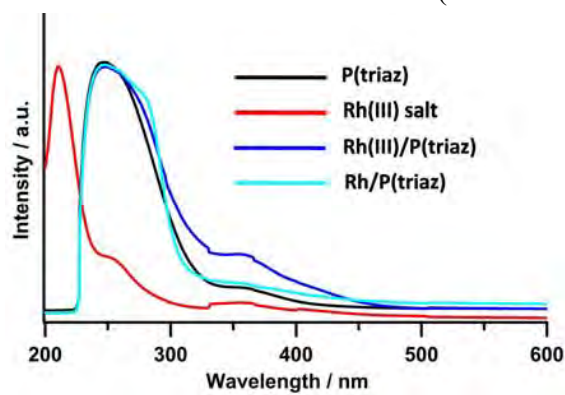

**Figure S33**. The UV-vis spectra monitoring the formation process of Rh/P(triaz) in dichloromethane-methanol mixture (volume ratio = 2:1).

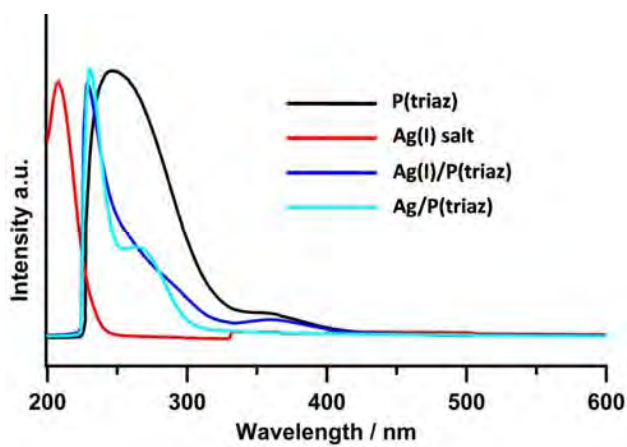

**Figure S34**. The UV-vis spectra monitoring the formation process of Ag/P(triaz) in dichloromethane-methanol mixture (volume ratio = 2:1).

S19

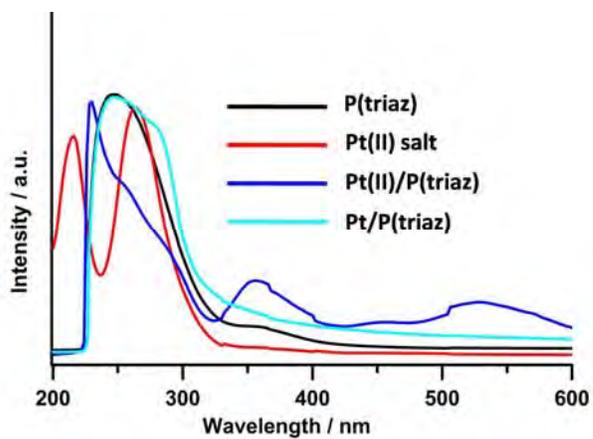

**Figure S35**. The UV-vis spectra monitoring the formation process of Pt/P(triaz) in dichloromethane-methanol mixture (volume ratio = 2:1).

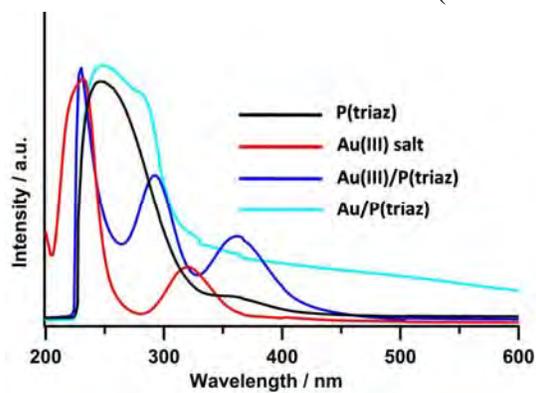

**Figure S36**. The UV-vis spectra monitoring the formation process of Au/P(triaz) in dichloromethane-methanol mixture (volume ratio = 2:1).

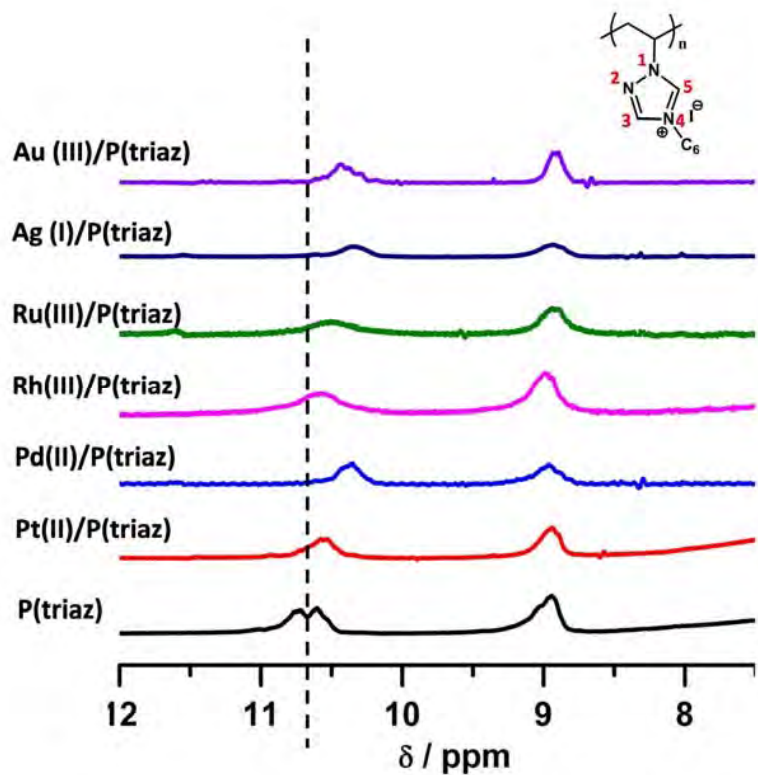

**Figure S37**. $^1$H NMR spectra of native P(triaz) and P(triaz) after mixing with different metal ion species (denoted as metal ion/P(triaz)) in $CD_2Cl_2$ and $CH_3OH$ (volume ratio = 2:1). An obvious shift of the H atom at C5 position (C-5 proton) to high magnetic field could be observed (shown in dot line) in metal ion/P(triaz) as compared with that of pure P(triaz), which is indicative of the coordination between the metal ions and P(triaz).

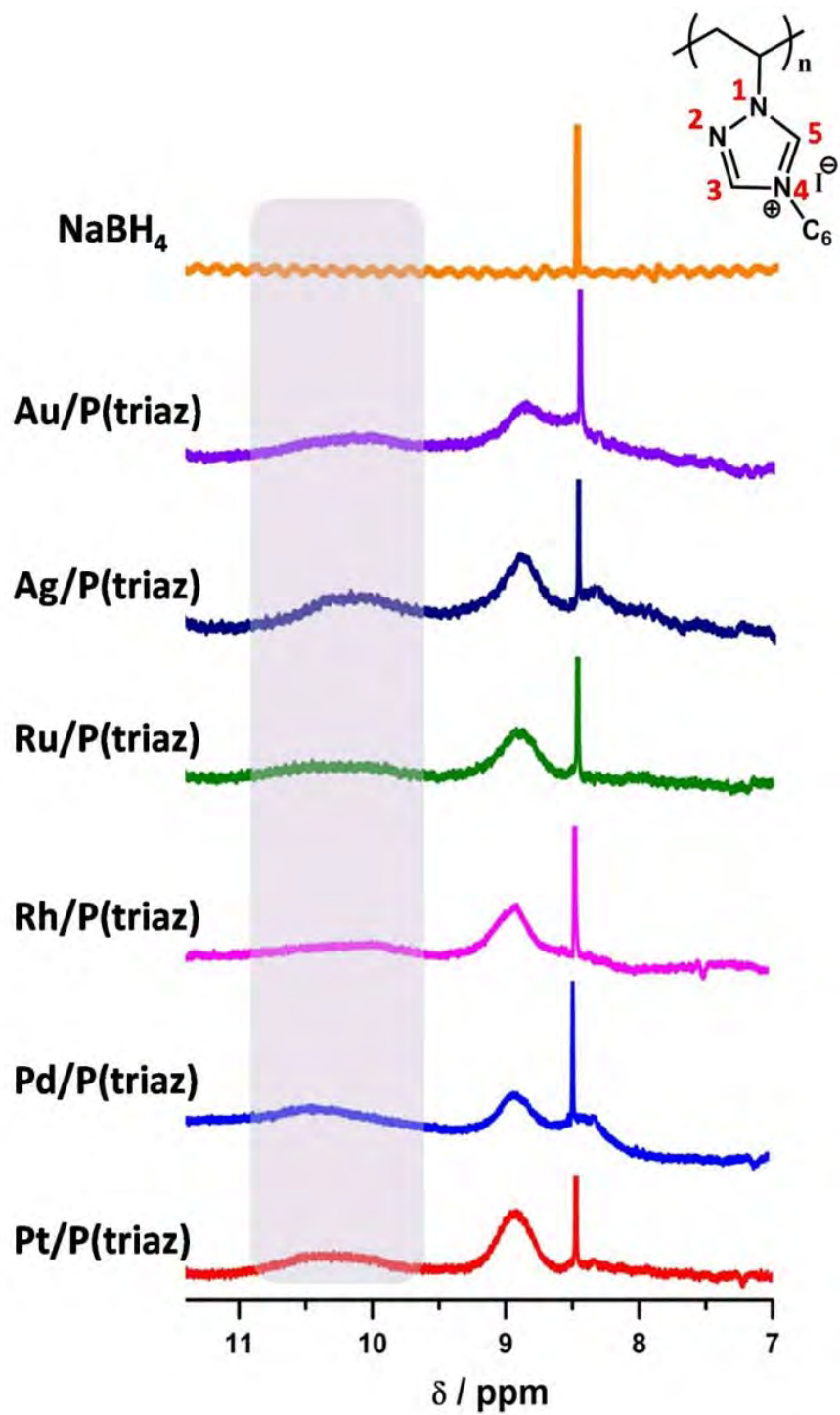

**Figure S38**. $^1$H NMR spectra of metal ion/P(triaz) after adding NaBH$_4$ (denoted as MC/P(triaz)) in CD$_2$Cl$_2$ and CH$_3$OH mixture (volume ratio = 2:1). The signal at 8.5 ppm in all spectra is attributed to the influence of the NaBH$_4$ as demonstrated by a control experiment via direct adding NaBH$_4$ into a CD$_2$Cl$_2$ and CH$_3$OH mixture (the yellow line in the top). The C-5 proton (10.6 ppm) in MC/P(triaz) decreased its intensity, indicating that P(triaz) *in situ* generates the polycarbene during the MC formation process.

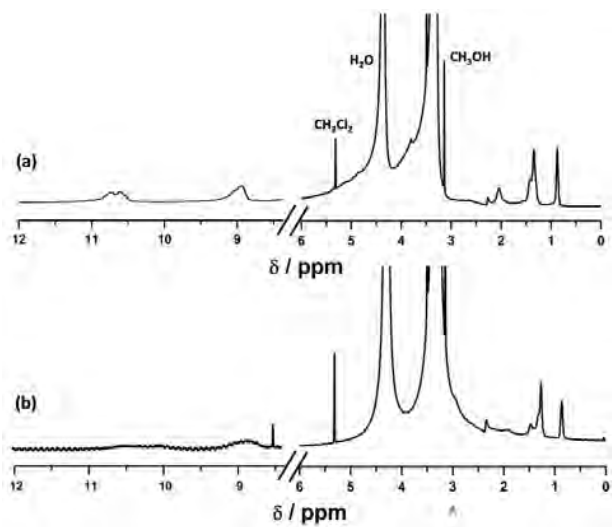

**Figure S39.** $^1$H-NMR spectra of P(triaz) in $CD_2Cl_2$ and $CH_3OH$ mixture (volume ratio = 2:1) before (a) and after (b) adding $NaBH_4$. The signal at 8.5 ppm in spectra (b) is attributed to $NaBH_4$ in the mixture $CD_2Cl_2$ and $CH_3OH$ (volume ratio = 2:1). The characteristic peaks maintain after adding $NaBH_4$ except the decreased intensity of C-5 protons. This indicates that the chemical structure of P(triaz) is stable in the MC formation process.

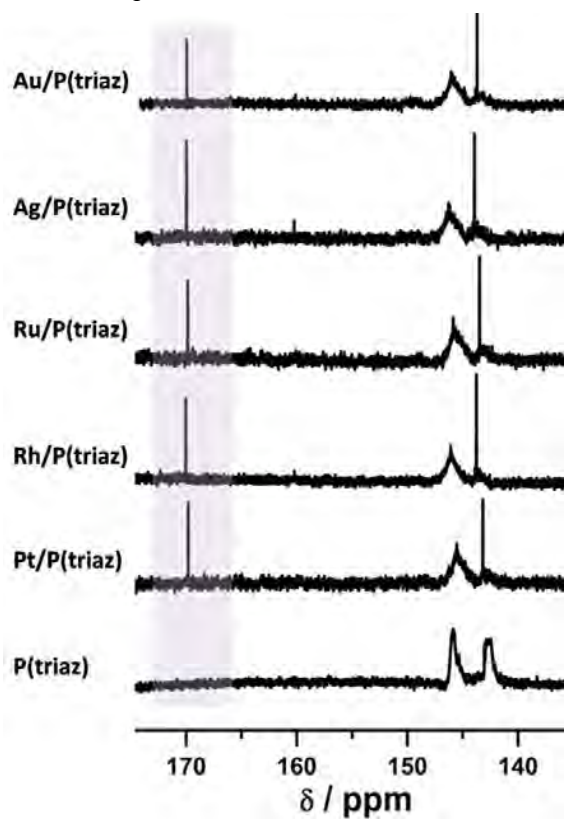

**Figure S40.** $^{13}$C NMR spectra of P(triaz), MCs/P(triaz) in $CD_2Cl_2$ and $CH_3OH$ mixture (volume ratio = 2:1). The strong intensive peaks at around 170 ppm appeared in MCs/P(triaz) (shown in violet rectangle), typical chemical shifts for metal-carbene coordination.

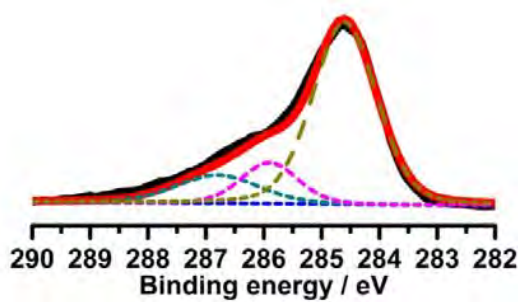

**Figure S41**. XPS spectra for C 1s signals of Co/P(triaz). The C1s spectra could be fitted by the sum of three separated peaks (dotted lines) with 1:1:8 area ratios that correspond to C5 (286.8 eV), C3 (285.9 eV) and eight alkane carbons (284.6 eV) in PIL. The C5 component shifts 0.4 eV to lower binding energy in Co/P(triaz) as compared with that of P(triaz) due to the Co-carbene complexation.

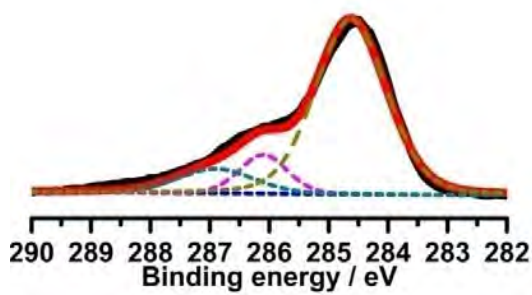

**Figure S42**. XPS spectra for C 1s signals of Ni/P(triaz). The C1s spectra could be fitted by the sum of three separated peaks (dotted lines) with 1:1:8 area ratios that correspond to C5 (286.9 eV), C3 (286.1 eV) and eight alkane carbons (284.6 eV) in PIL. The C5 component shifts 0.3 eV to lower binding energy in Ni/P(triaz) as compared with that of P(triaz) due to the Ni-carbene complexation.

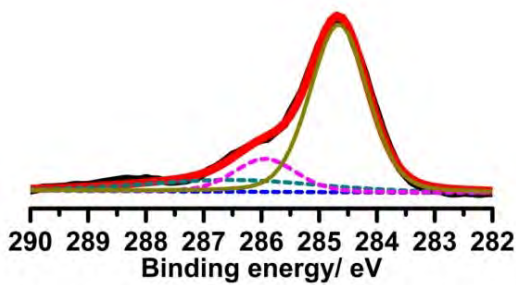

**Figure S43**. XPS spectra for C 1s signals of Cu/P(triaz). The C1s spectra could be fitted by the sum of three separated peaks (dotted lines) with 1:1:8 area ratios that correspond to C5 (286.7 eV), C3 (286.0 eV) and eight alkane carbons (284.6 eV) in PIL. The C5 component shifts 0.5 eV to lower binding energy in Cu/P(triaz) as compared with that of P(triaz) due to the Cu-carbene complexation.

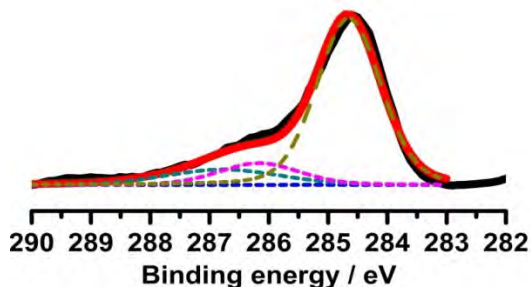

**Figure S44**. XPS spectra for C 1s signals of Ru/P(triaz). The C1s spectra could be fitted by the sum of three separated peaks (dotted lines) with 1:1:8 area ratios that correspond to C5 (286.8 eV), C3 (286.1 eV) and eight alkane carbons (284.6 eV) in PIL. The C5 component shifts 0.4 eV to lower binding energy in Ru/P(triaz) as compared with that of P(triaz) due to the Ru-carbene complexation.

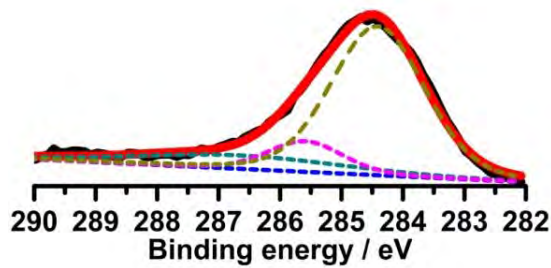

**Figure S45**. XPS spectra for C 1s signals of Rh/P(triaz). The C1s spectra could be fitted by the sum of three separated peaks (dotted lines) with 1:1:8 area ratios that correspond to C5 (286.6 eV), C3 (286.1 eV) and eight alkane carbons (284.6 eV) in PIL. The C5 component shifts 0.6 eV to lower binding energy in Rh/P(triaz) as compared with that of P(triaz) due to the Rh-carbene complexation.

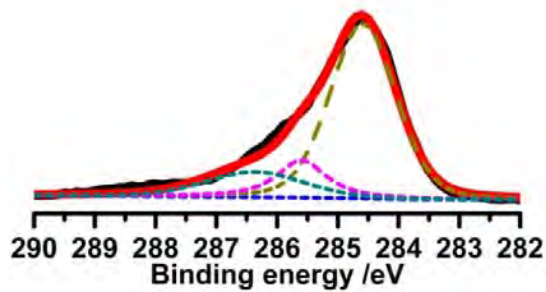

**Figure S46**. XPS spectra for C 1s signals of Ag/P(triaz). The C1s spectra could be fitted by the sum of three separated peaks (dotted lines) with 1:1:8 area ratios that correspond to C5 (286.4 eV), C3 (285.6 eV) and eight alkane carbons (284.6 eV) in PIL. The C5 component shifts 0.8 eV to lower binding energy in Ag/P(triaz) as compared with that of P(triaz) due to the Ag-carbene complexation.

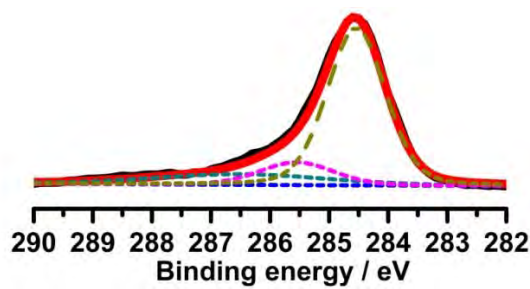

**Figure S47**. XPS spectra for C 1s signals of Pt/P(triaz). The C1s spectra could be fitted by the sum of three separated peaks (dotted lines) with 1:1:8 area ratios that correspond to C5 (286.8 eV), C3 (285.6 eV) and eight alkane carbons (284.6 eV) in PIL. The C5 component shifts 0.4 eV to lower binding energy in Pt/P(triaz) as compared with that of P(triaz) due to the Pt-carbene complexation.

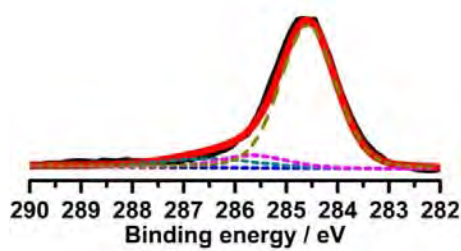

**Figure S48**. XPS spectra for C 1s signals of Au/P(triaz). The C1s spectra could be fitted by the sum of three separated peaks (dotted lines) with 1:1:8 area ratios that correspond to C5 (286.4 eV), C3 (285.6 eV) and eight alkane carbons (284.6 eV) in PIL. The C5 component shifts 0.8 eV to lower binding energy in Au/P(triaz) as compared with that of P(triaz) due to the Au-carbene complexation.

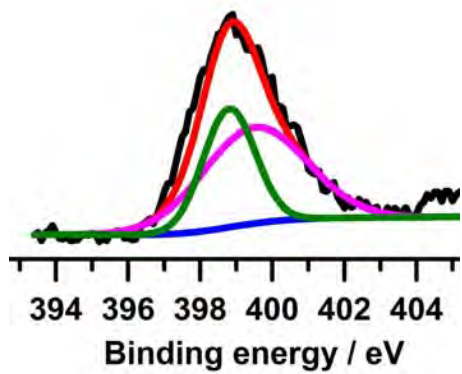

**Figure S49**. XPS spectra for N 1s signals of Co/P(triaz). N1s spectra could be fitted by the sum of two separate peaks (at 398.8 and 399.6 eV) with a ratio of 1:2 in their integration that correspond to naked nitrogen (N2) and the two other nitrogen atoms (N1 and N4) of the triazolium ring, respectively. Compared with the P(triaz) (N2 at 398.6 eV) (Figure 3d in main text), a shift of binding energy of N2 toward high position could be observed in Co/P(triaz), which could be attributed to the resultant binding to the metal.

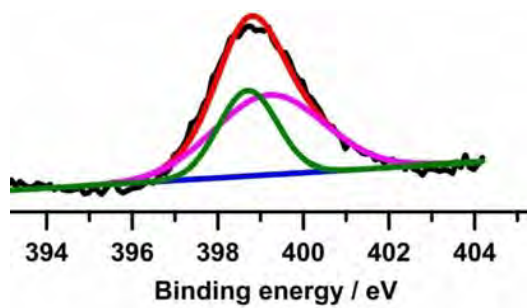

**Figure S50**. XPS spectra for N 1s signals of Ni/P(triaz). XPS spectra for N 1s signals of Ni/P(triaz). N1s spectra could be fitted by the sum of two separate peaks (at 398.8 and 399.2 eV) with a ratio of 1:2 in their integration area that correspond to naked nitrogen (N2) and the two other nitrogen atoms (N1 and N4) of the triazolium ring, respectively. Compared with the P(triaz) (N2 at 398.6 eV), a shift of binding energy of N2 toward high position could be observed in Ni/P(triaz), which could be attributed to the resultant binding to the metal.

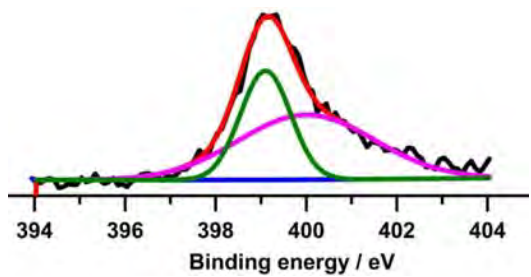

**Figure S51**. XPS spectra for N 1s signals of Cu/P(triaz). N1s spectra could be fitted by the sum of two separate peaks (at 399.1 and 400 eV) with a ratio of 1:2 in their integration area that correspond to naked nitrogen (N2) and the two other nitrogen atoms (N1 and N4) of the triazolium ring, respectively. Compared with the P(triaz) (N2 at 398.6 eV), a shift of binding energy of N2 toward high position could be observed in Cu/P(triaz), which could be attributed to the resultant binding to the metal.

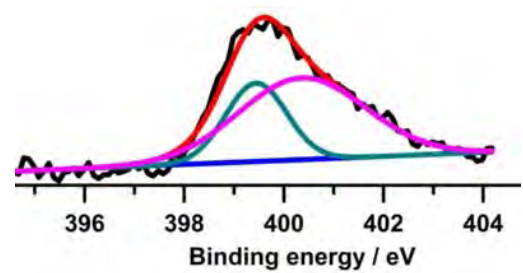

**Figure S52**. XPS spectra for N 1s signals of Ru/P(triaz). N1s spectra could be fitted by the sum of two separate peaks (at 399.5 and 400.4 eV) with a ratio of 1:2 in their integration area that correspond to naked nitrogen (N2) and the two other nitrogen atoms (N1 and N4) of the triazolium ring, respectively. Compared with the P(triaz) (N2 at 398.6 eV), a shift of binding energy of N2 toward high position could be observed in Ru/P(triaz), which could be attributed to the resultant binding to the metal.

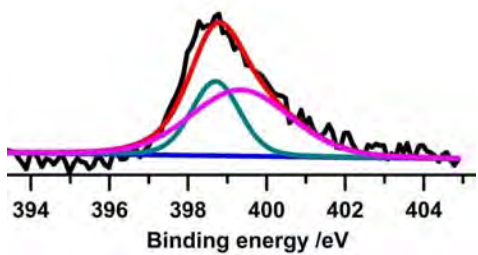

**Figure S53**. XPS spectra for N 1s signals of Rh/P(triaz). N1s spectra could be fitted by the sum of two separate peaks (at 398.7 and 399.4 eV) with a ratio of 1:2 in their integration area that correspond to naked nitrogen (N2) and the two other nitrogen atoms (N1 and N4) of the triazolium ring, respectively. Compared with the P(triaz) (N2 at 398.6 eV), a shift of binding energy of N2 toward high position could be observed in Rh/P(triaz), which could be attributed to the resultant binding to the metal.

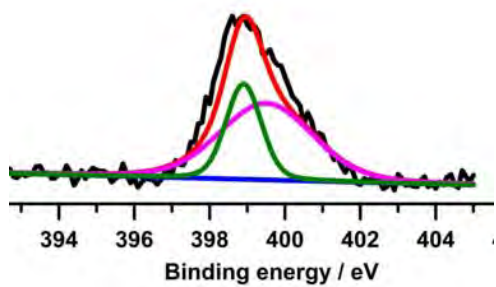

**Figure S54**. XPS spectra for N 1s signals of Ag/P(triaz). N1s spectra could be fitted by the sum of two separate peaks (at 398.9 and 399.5 eV) with a ratio of 1:2 in their integration area that correspond to naked nitrogen (N2) and the two other nitrogen atoms (N1 and N4) of the triazolium ring, respectively. Compared with the P(triaz) (N2 at 398.6 eV), a shift of binding energy of N2 toward high position could be observed in Ag/P(triaz), which could be attributed to the resultant binding to the metal.

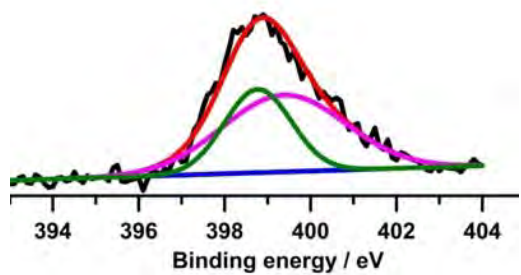

**Figure S55**. XPS spectra for N 1s signals of Pt/P(triaz). N1s spectra could be fitted by the sum of two separate peaks (at 398.8 and 399.4 eV) with a ratio of 1:2 in their integration area that correspond to naked nitrogen (N2) and the two other nitrogen atoms (N1 and N4) of the triazolium ring, respectively. Compared with the P(triaz) (N2 at 398.6 eV), a shift of binding energy of N2 toward high position could be observed in Pt/P(triaz), which could be attributed to the resultant binding to the metal.

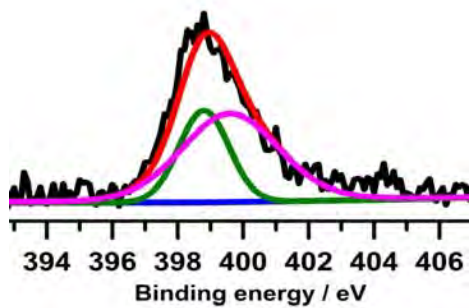

**Figure S56**. XPS spectra for N 1s signals of Au/P(triaz). N1s spectra could be fitted by the sum of two separate peaks (at 398.8 and 399.6 eV) with a ratio of 1:2 in their integration area that correspond to naked nitrogen (N2) and the two other nitrogen atoms (N1 and N4) of the triazolium ring, respectively. Compared with the P(triaz) (N2 at 398.6 eV), a shift of binding energy of N2 toward high position could be observed in Au/P(triaz), which could be attributed to the resultant binding to the metal.

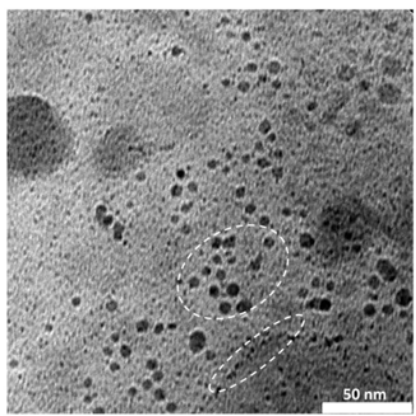

**Figure S57**. The bright field (BF) TEM image of Pd/PIL-imidaz. The white circles indicate that both clusters and large nanoparticles exist.

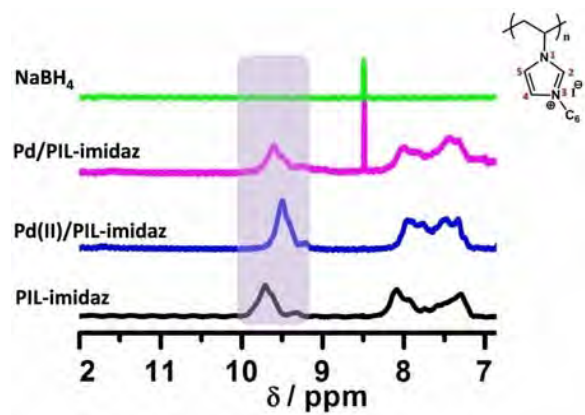

**Figure S58.** The $^1$H NMR spectra recorded the formation process of Pd/PIL-imidaz in a $CD_2Cl_2$ and $CH_3OH$ mixture (volume ratio = 2:1). The signal of H atom at C2 position (9.7 ppm) in native PIL-imidaz remains in Pd/PIL-imidaz after addition of $NaBH_4$ (shown in violet rectangle), indicating that the formation of carbene carbon in PIL-imidaz is much less than that of P(triaz) in the presence of NaBH$_4$.

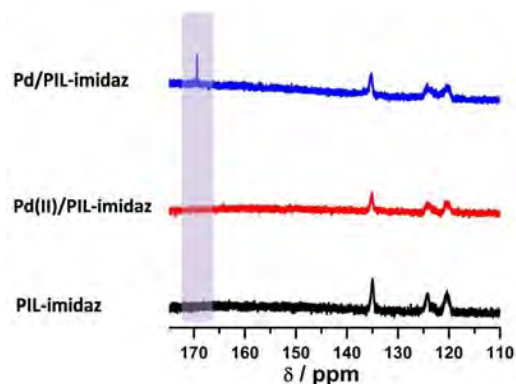

**Figure S59**. $^{13}$C NMR spectra of PIL-imidaz, Pd(II)/PIL-imidaz and Pd/PIL-imidaz in CD$_2$Cl$_2$ and CH$_3$OH mixture (volume ratio = 2:1). In comparison to the PIL-imidaz, only a weak peak at 169.6 ppm appeared in Pd/PIL-imidaz (shown in violet rectangle). This phenomenon indicates that the PIL-imidaz is less efficient to generate polycarbene to control cluster formation in compared to P(triaz) after treated by NaBH$_4$. This is also consistent with the observation in the TEM image of non-uniform distribution of particles of Pd/PIL-imidaz (Figure S57).

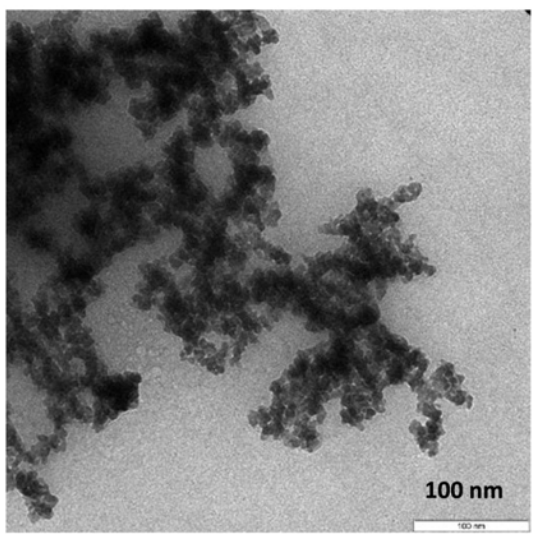

**Figure S60**. Bright field (BF) TEM image of the Pd/(triazolium monomer) formed by using triazolium monomer as stabilizer.

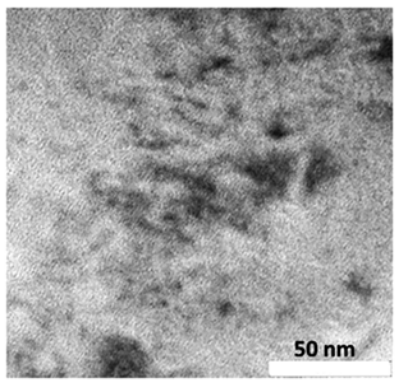

**Figure S61**. The TEM images of the Ag/P(triaz) nanoparticles prepared by using UV light radiation as reducing source.

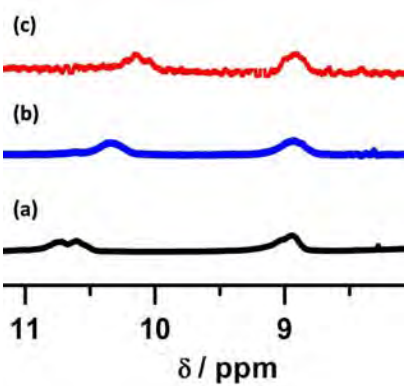

**Figure S62**. The $^1$H NMR spectra of (a) P(triaz), (b) Ag(I)/P(triaz) and (c) Ag(I)/P(triaz) after reduction by UV irradiation in a $CD_2Cl_2$ and $CH_3OH$ mixture (volume ratio = 2:1). The proton at C5 position has little-to-no change in its intensity, indicating that no detectable amount of carbenes was generated in this photoreduction process.

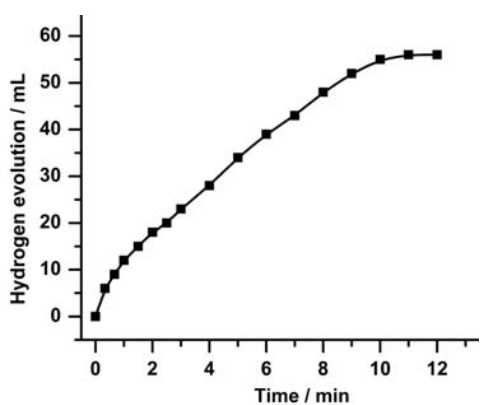

**Figure S63**. Time course plot of $H_2$ generation for the methanolysis of AB over the commercial Pd/C catalyst at 298 K (Pd/AB =0.01) (TOF: 22.4 min$^{-1}$).

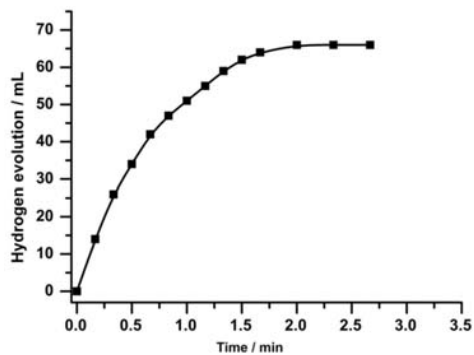

**Figure S64**. Time course plot of H$_2$ generation for the methanolysis of AB over the Rh/ triazolium monomer catalyst at 298 K (Rh/AB =0.01).

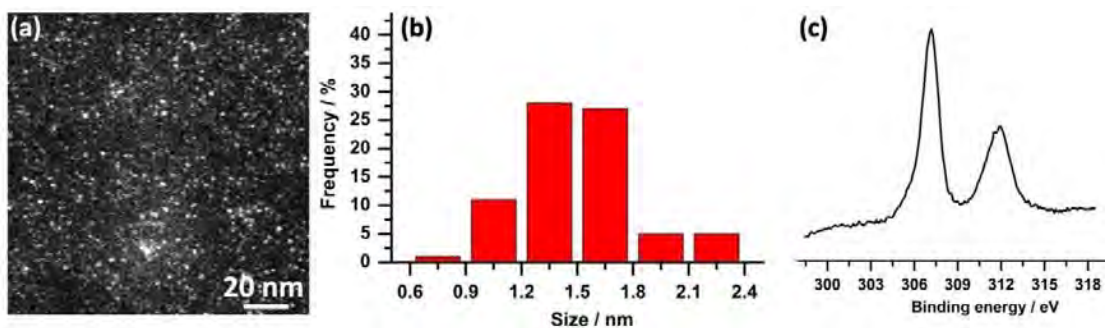

**Figure S65**. (a) HAADF-STEM image of Rh/PAMAM-OH catalyst and (b) the corresponding size distribution histogram of Rh clusters. (c) The XPS spectrum of Rh cluster.

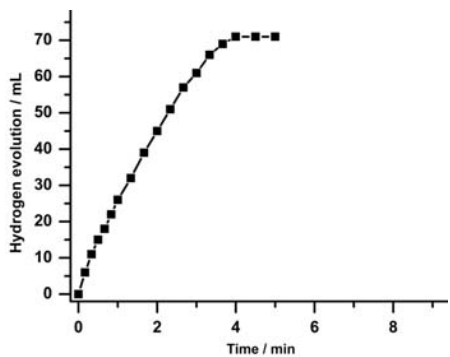

**Figure S66.** Time course plot of H$_2$ generation for the methanolysis of AB over the Rh/PAMAM-OH catalyst at 298 K (Rh/AB =0.01).

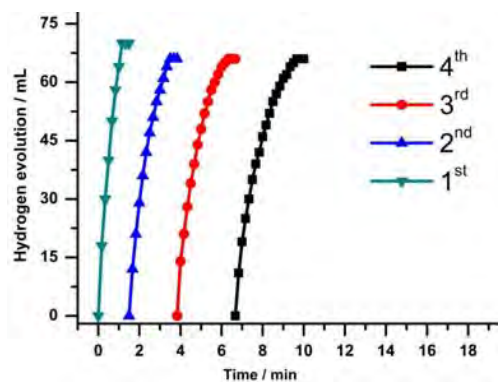

**Figure S67**. Durability test for AB methanolysis reaction over Rh/P(triaz) catalyst at 298 K (Rh/AB = 0.01).

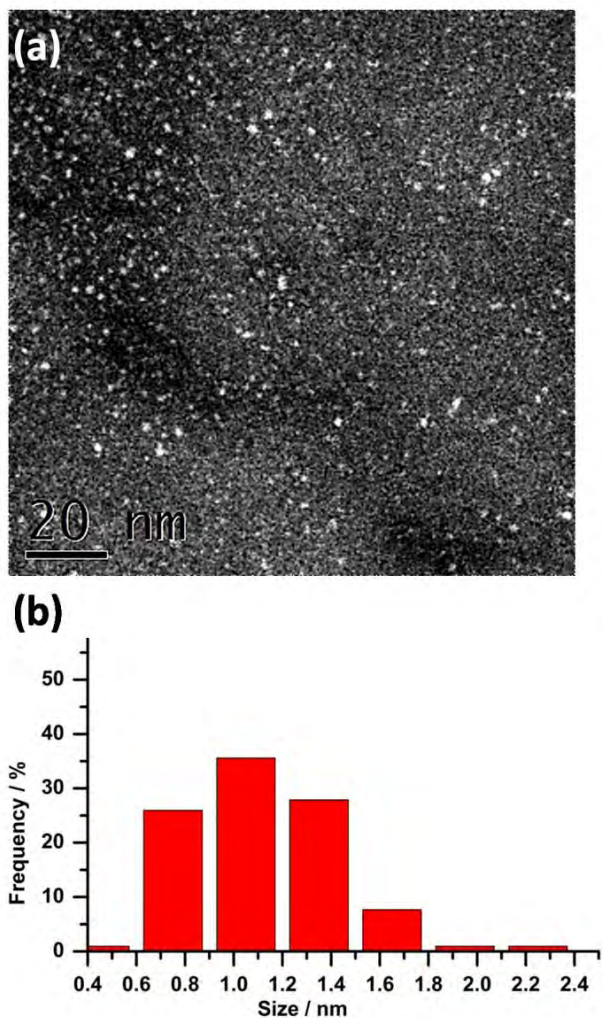

**Figure S68**. (a) STEM-HADDF image and (b) the size distribution histogram of Rh cluster of the as-synthesized Rh/P(triaz) catalyst before AB methanolysis reaction.

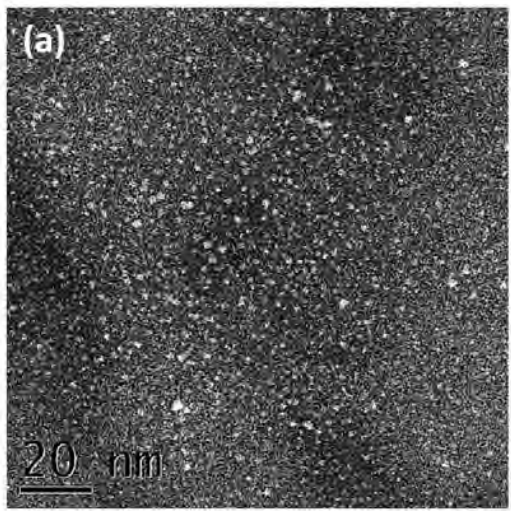

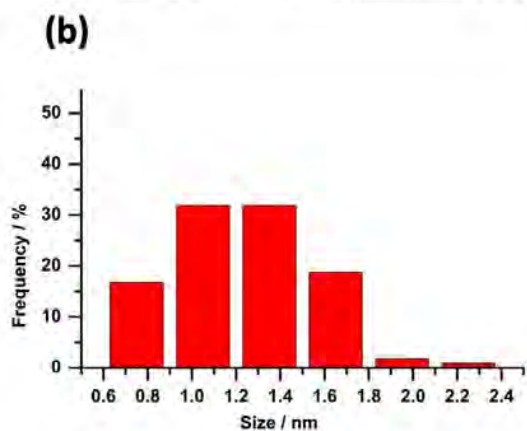

**Figure S69**. (a) STEM-HADDF image and (b) the size distribution histogram of Rh cluster of the Rh/P(triaz) catalyst after AB methanolysis reaction (Rh/AB = 0.01).

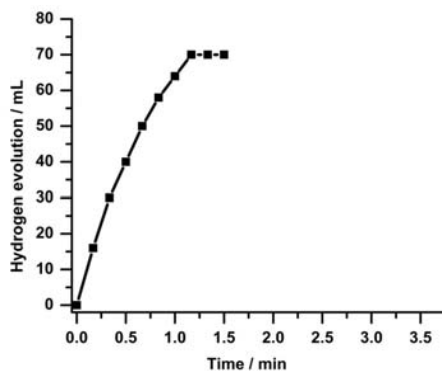

**Figure S70**. Time course plot of $H_2$ generation for the methanolysis of AB over the recycling catalyst generated by redissolving the $N_2$-dried Rh/P(triaz) catalyst (Rh/AB = 0.01) in a dichloromethane and methanol mixture (volume ratio = 2:1).

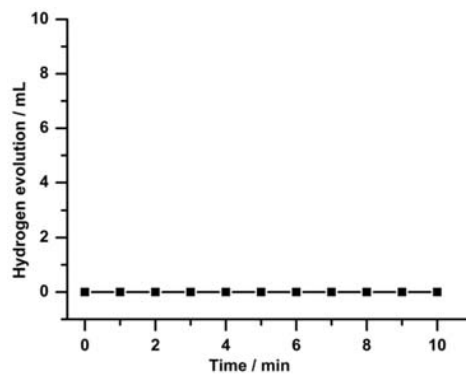

**Figure S71.** Time course plot of H$_2$ generation for the methanolysis of AB over only P(triaz).

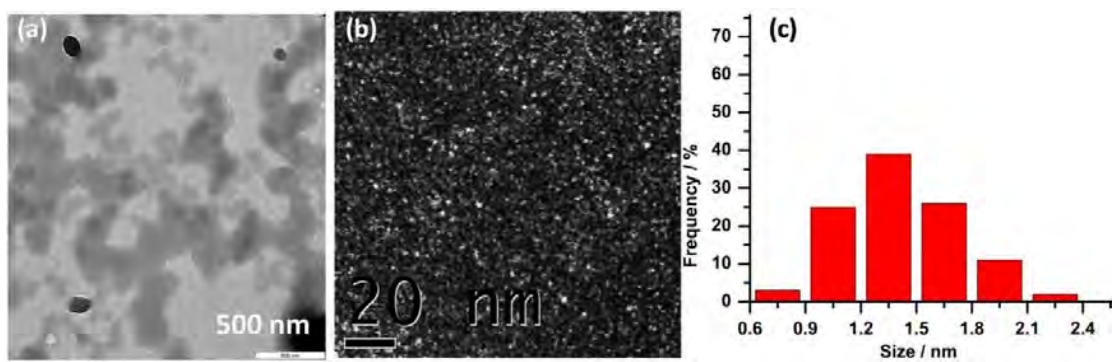

**Figure S72.** (a) TEM image of P(triaz) and (b) HAADF-STEM image of Rh/P(triaz) dried from their methanol solutions. (c) The corresponding size distribution histogram of Rh clusters (1.4 ± 0.3 nm).

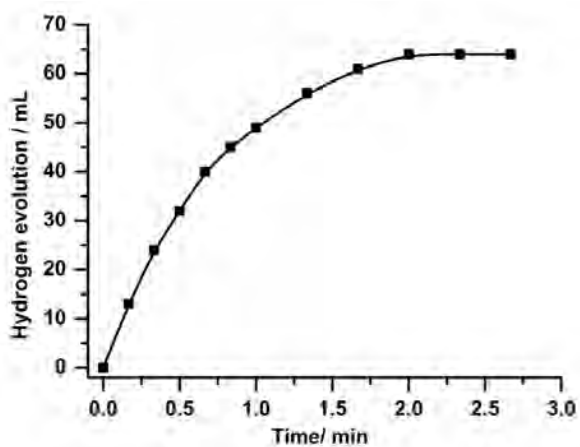

**Figure S73.** Time course plot of H$_2$ generation for the methanolysis of AB over the Rh/P(triaz) catalyst (Rh/AB = 0.01) in methanol.

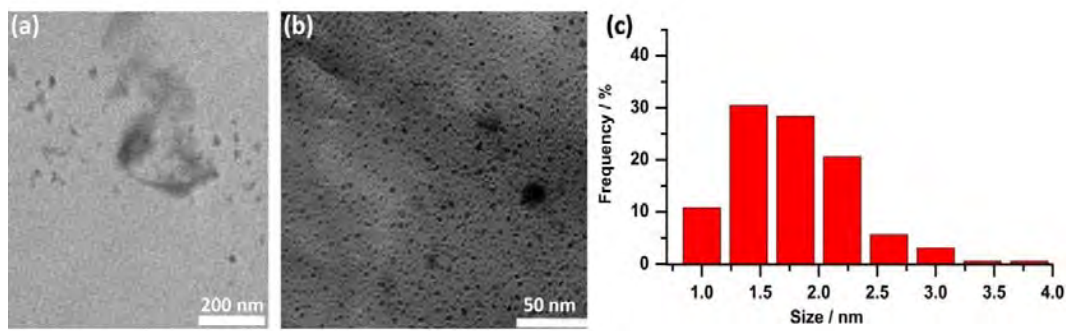

**Figure S74.** TEM image of (a) PIL-butyl and (b) Rh/PIL-butyl dried from their in dichloromethane and methanol mixture (volume ratio = 2:1). (c) The corresponding size distribution histogram of Rh clusters (1.8 ± 0.4 nm).

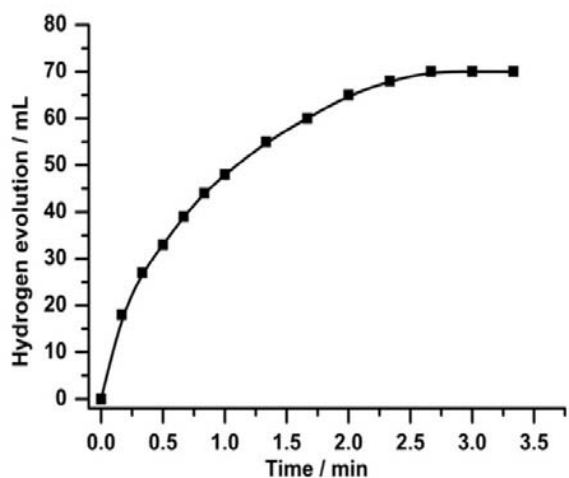

**Figure S75.** Time course plot of $H_2$ generation for the methanolysis of AB over the Rh/PIL-butyl catalyst (Rh/AB = 0.01) in a dichloromethane and methanol mixture (volume ratio = 2:1).

**Table S1. The calculated molar ratio of carbene/metal in the meetal/P(triaz) products.**

|    | Amount of metal (mmol)[a] | Amount of carbene generated in P(triazo) (mmol)[b] | Carbene/metal molar ratio |
|----|---------------------------|----------------------------------------------------|---------------------------|
| Ru | 0.01                      | 0.0111                                             | 1.11                      |
| Rh | 0.0049                    | 0.0098                                             | 2.00                      |
| Pd | 0.0047                    | 0.0054                                             | 1.15                      |
| Ag | 0.0046                    | 0.0046                                             | 1.00                      |
| Pt | 0.0026                    | 0.0047                                             | 1.81                      |
| Au | 0.0025                    | 0.0051                                             | 2.04                      |

[a] The metal content in each specie is 0.5 mg except 1 mg for Ru.
[b] The amount of carbene is calculated by integrating the C-5 proton signal in the $^1$H NMR spectra in Figure S38 and comparing this value with native P(triaz). The origin amount of P(triaz) is 5 mg for each metal specie.